\documentclass[10pt]{IEEEtran}
\usepackage[utf8]{inputenc}
\usepackage{textcomp}

\usepackage{cancel}
\usepackage{graphicx}
\PassOptionsToPackage{final}{graphicx}
\usepackage{amsmath}
\usepackage{amssymb}

\usepackage{amsthm}
\usepackage[dvipsnames]{xcolor}
\usepackage[version=4]{mhchem}
\usepackage{siunitx}
\usepackage{longtable,tabularx}
\usepackage{mathtools}
\usepackage{easyReview}
\usepackage{cleveref}
\newcommand{\diag}{\textrm{diag}}

\newcommand{\norm}[1]{\left\lVert#1\right\rVert}
\newcommand{\gap}{,\quad\quad}
\newcommand{\mathtext}[1]{\textrm{#1}}

\newcommand{\dif}{\ensuremath{\; \textrm{d}}}
\newcommand{\R}{\mathbb{R}}
\newcommand{\Gr}{\ensuremath{\mathcal{G}}}
\newcommand{\Lap}{\ensuremath{\mathcal{L}}}
\newcommand{\Adj}{\ensuremath{\mathcal{A}}}
\newcommand{\Lapu}{\ensuremath{\mathcal{Q}}}
\newcommand{\Deg}{\ensuremath{\mathcal{D}}}
\newcommand{\one}{\ensuremath{\mathbf{1}}}

\newtheorem{thm}{Theorem}[section]
\newtheorem{lem}{Lemma}[section]
\newtheorem{prop}{Proposition}[section]

\theoremstyle{definition}
\newtheorem{fact}{Fact}[section]
\newtheorem{defn}{Definition}
\newtheorem{assum}{Assumption}

\theoremstyle{remark}
\newtheorem{rem}{Remark}[section]

\begin{document}

\title{Consensus of networked double integrator systems under
sensor bias}

\author{Pallavi Sinha, Srikant Sukumar and Himani Sinhmar%
\thanks{S. Srikant (\textit{srikant.sukumar@iitb.ac.in}) and P. Sinha (\textit{pallavisinha33@gmail.com}) are with the Indian Institute of Technology Bombay, Mumbai, India and H. Sinhmar (\textit{himani.sinhmar@gmail.com}) is with Cornell University, Ithaca, USA}}

\maketitle

\begin{abstract}
 A novel distributed control law for consensus of networked double integrator systems with biased measurements is developed in this article. The agents measure relative positions over a time-varying, undirected graph with an unknown and constant sensor bias corrupting the measurements. An adaptive control law is derived using Lyapunov methods to estimate the individual sensor biases accurately. The proposed algorithm ensures that position consensus is achieved exponentially in addition to bias estimation. The results leverage recent advances in collective initial excitation based results in adaptive estimation. Conditions connecting bipartite graphs and collective initial excitation are also developed. The algorithms are illustrated via simulation studies on a network of double integrators with local communication and biased measurements.
\end{abstract}


% \newpage
% \begin{frontmatter}

% 	\runtitle{Consensus of networked E-L systems under measurement bias}

% 	\title{Consensus of networked Euler-Lagrange systems under measurement bias\thanksref{footnoteinfo}}
% 	\thanks[footnoteinfo]{%
% 	Emails: \texttt{himani.sinhmar@gmail.com, srikant.sukumar@iitb.ac.in}%
% }
% 	\author[HS]{Himani Sinhmar}
% 	\author[SS]{Sukumar Srikant}

% 	\address[HS]{Department of Aerospace Engineering, Indian Institute of Technology Bombay, Powai, Mumbai 400076, India.}
% 	\address[SS]{Systems \& Control Engineering, Indian Institute of Technology Bombay, Powai, Mumbai 400076, India.}

% 	\begin{keyword}
% 		sparse control, networked control, optimal control
% 	\end{keyword}

% 	%\date{\DTMnow}

% 	\begin{abstract}
% A novel distributed control law for consensus of networked Euler-Lagrange systems with biased measurements is developed in this article. The agents measure relative positions with a non-zero, unknown and constant sensor bias corrupting the measurements. An adaptive control law is derived using Lyapunov methods to accurately estimate the individual sensor biases. The proposed algorithm ensures that the trajectories converge to that of a leader exponentially in addition to the bias estimation. The results leverage recent advances in initial excitation based results in adaptive estimation. The algorithms are illustrated via simulation studies on a network of two-link, revolute-joint arms tracking a leader with only local communication and measurements.
% 	\end{abstract}

% \end{frontmatter}

\section{Introduction}
% In recent years there has been a growing interest in the field of swarm robotics. Advances in communication and low-cost sensor technology have enabled cooperative multi-agent systems to have many practical applications. In the distributed control paradigm, each agent bases its actions on information derived from immediate neighbors, allowing the agents to execute a control law without requiring information of the network as a whole. A distributed control naturally improves scalability and robustness to system failure. Spacecraft rendezvous, rigid body attitude synchronization and formation flight are examples of scenarios where the agents need to update their (information) states to reach a consensus \cite{slotine_sff}, \cite{weiren_application}. 
Consensus of networked double integrators has been studied extensively in control literature and several globally convergent controllers have been proposed \cite{qin2011second,xie2017global,wang2008finite,meng2013global,wang2013global}. This degree of interest is because the double integrator is one of the most fundamental block in any control system. Applications of double integrators include feedback linearizable nonlinear mechanical and aerospace systems such as free-rigid body motion, manipulator motion and spacecraft rotation. The consensus algorithms thus obtained can be further extended to complex nonlinear systems.
Multi-agent systems with double integrator dynamics have been extensively studied in literature. See for example \cite{weiren} and references therein. A comprehensive survey of the several consensus results in literature can be found in \cite{AA21-consensusrev}. The preceding references, however, assume perfect measurements or extraneous disturbances only.

 The motivation for the problem of consensus under sensor bias originates from mechanical systems that have
only relative position and absolute velocity measurements available
for feedback. However, relative position sensors suffer from
errors such as bias in measurements. Unknown biases can appear during the functioning of various sensors such as rate gyros, accelerometers, magnetometers, altimeters, range sensors etc. These biases can be an outcome of inaccurate sensor calibration, environmental conditions, etc. The presence of bias deteriorates the performance of control laws on the network, and may result in stability issues \cite{liu2016robust,meng2016formation,de2014controlling}. Specifically, bias in relative position feedback could drive the agents to infinity, if not compensated. It is, therefore, of interest to estimate the biases and possibly nullify their effect on the network.
In the context of the continuous system, bias uncertainties in
measurements are in general, sparsely studied. In the context
of a single rigid-body system, ‘gyro bias’ is the most
commonly addressed bias uncertainty and has been studied
in detail in several references, including \cite{thienel2003coupled,boskovic2000globally,metni2005attitude}.
However, the literature on adaptive estimation and compensation of `position'
sensor bias is somewhat limited and
the only relevant contributions known to the authors are by \cite{mrac,troni2013adaptive}. There have of course been parallel approaches using non-smooth control laws, where disturbance rejection is possible for both single and networked second-order systems subject to knowledge of bounds on the bias uncertainty which is then modeled as a bounded disturbance. Such non smooth laws for disturbance rejection in double integrator systems have been explored in \cite{slidingmode2,slidingmode1,ext_dist,nl_wang}. An approach to estimating measurement inconsistencies using an output regulation-based technique is presented in \cite{DCJ-15}. Accurate estimation in \cite{DCJ-15}, however requires a unique constant, graph structure.
%\par For uncertain double integrator networked systems (sensor bias being one such uncertainty), \cite{mahyuddin2012finite} proposes a novel adaptive control algorithm using a finite time adaptive estimator with local agent's information.
\par For uncertain networked double integrators (sensor bias being one such uncertainty), conventional adaptive control laws (including \cite{narendra2012stable,boyd1983parameter,mahyuddin2012finite,sun2016adaptive,cpechen2013distributed}) require the regressor function to be persistently exciting (PE) or collectively persistently exciting(C-PE) for parameter convergence. Recently, several methods have been proposed to get rid of the PE condition for parameter convergence. \cite{akella} proposes an adaptive algorithm that uses both instantaneous state data and past measurements for the adaptation process. This scheme ensures parameter estimation errors converge to zero exponentially, subject to the satisfaction of a finite-time excitation condition. In the same spirit, \cite{sbasu} proposes a PI-like (Proportional-Integral controllers) parameter update law that guarantees parameter convergence with a relaxation of the PE condition, namely \emph{initial excitation} (IE) on the regressor. \cite{garg2020distributed} is an extension of \cite{sbasu}, where the authors develop a distributed composite adaptive synchronization algorithm for multiple uncertain Euler-Lagrange (EL) systems to ensure parameter convergence using the \emph{collective-IE (C-IE)} condition. The method proposed in \cite{sbasu} obviates the need for data-storage and memory allocation required in concurrent learning-based adaptive control~\cite{cymj13} methods. 

% One approach to bias rejection is to use disturbance rejection schemes. In this context, a non-smooth distributed model-independent algorithm for a directed and switching network of Euler-Lagrange agents to track the trajectory of a leader in the presence of bounded disturbance is addressed in \cite{anderson}. A second-order super-twisting sliding mode controller is developed in \cite{slidingmode2} for consensus tracking of a leader-follower network in the presence of uncertainties and external disturbances. However, the approach in \cite{slidingmode2} can ensure finite-time consensus only if the bounds on the disturbances and parameter uncertainties are known to the agents. In \cite{slidingmode1} a continuous sliding mode tracking protocol with an adaptive mechanism is developed for the consensus tracking problem of second-order nonlinear networked system with known bounds on disturbances and actuator faults. The coordination control problem of a network of heterogeneous first and second order systems with known bounds on external disturbance is also investigated in \cite{ext_dist}. A composite consensus control strategy is proposed for a second-order multi-agent system with the assumption of known bounds on mismatched disturbances in \cite{nl_wang}. All the aforementioned disturbance rejection based approaches rely on non-smooth control laws and require explicit information of disturbance bounds.

There have been several strides in consensus under bounded disturbance and zero-mean noise. In \cite{wo_bound}, a leader-follower consensus control for a network of double integrators is proposed for follower measurements corrupted by (zero-mean) noise. This control law ensures that consensus tracking is achieved in the mean square sense for both fixed and switching communication networks. However, no bias errors are accounted for in this work. \cite{R2kingston2005consensus} shows a Kalman filter inspired technique for consensus which is input to state stable. As would be expected, the accuracy of the cooperation objective is directly related to the power level of the communication noise. \cite{R3garulli2011analysis} analyzes the asymptotic properties of linear consensus algorithms in the presence of bounded measurement errors. Here, consensus is not guaranteed with respect to all possible noise realizations. In \cite{R4shi2019self}, a novel self-triggering co-ordination scheme for finite time consensus is proposed in the presence of unknown but bounded noise affecting the communication channels. Bias errors with unknown bounds are not the subject of study in any of the above articles. 

We now summarize some of the results that lead up to the current work on consensus under measurement bias with unknown bounds. \cite{tandon} proposes an adaptive control law in the presence of unknown constant bias for a double integrator network. This controller ensures bounded closed-loop signals in the presence of sensor bias which would not be the case in the absence of adaptation. However, convergence only to a neighborhood of consensus can be shown. \cite{tandon} is based on the results in \cite{mrac} which addresses the problem of accommodating unknown sensor bias in a direct MRAC setting for a single agent. In \cite{loria}, the authors 
present a consensus algorithm for synchronization of double integrators over directed graphs in the presence of constant bias with unknown bounds. Here, the authors assume the existence of a bias error on each communication channel. Similar to \cite{tandon}, here too convergence `near' a common equilibrium point is guaranteed. \cite{himani} shows an extension of \cite{tandon,loria} to develop a distributed consensus tracking algorithm for spacecraft in formation modeled as an Euler-Lagrange network with similar bounded performance results. For a \emph{fixed} communication graph, exact constant estimation of a constant bias and consensus in single integrator agents are demonstrated in \cite{SDTM-20}. An undirected, connected, and non-bipartite graph network is shown to be necessary and sufficient for estimation of the full bias vector. 

In comparison with existing literature, the novel contributions of this article are as below.

\begin{itemize}
    \item Adaptive control laws for exact bias estimation and consensus are developed over \cite{tandon,himani,loria}. One sensor attached to each node is considered and \emph{all} relative measurements from a node are assumed to be affected by the same bias in contrast to \cite{loria}.
    %\item A nonlinear Euler Lagrange network is considered here over \cite{SDTM-20,tandon,loria}.
    \item A time-varying communication network topology is considered over \cite{SDTM-20,tandon,loria,himani}. The analysis is based on jointly connected and jointly non-bipartite graphs.
    \item A non-bipartite property of the communication graph for a finite initial-time only is shown to be necessary in contrast to \cite{SDTM-20}, where a constant non-bipartite graph is assumed for all time.
    \item A collective initial excitation based adaptive controller is employed for the first time in bias estimation problems over networks.
\end{itemize}

This paper is organized as follows. Section II introduces mathematical notation, necessary lemmas and in brief, graph theory. In section III, we formulate the consensus problem over a network of double integrator systems. We develop an adaptive control law for achieving consensus and bias estimation in section IV. A discussion on the choice of control gains and the collective initial excitation condition on the regressor matrix are presented in section V. Section VI presents numerical simulations validating our algorithm. Conclusions are presented in section VII.

\section{ Preliminaries}

In this section, we present several mathematical notations, lemmas, assumptions, and a concise introduction of graph theory that forms the basis of the problem formulation.

\subsection{Notation}

% The signum function, $\sgn(\cdot)$, for a vector $x = [x_1 \cdots x_n]^T$ $\in$ $\mathbb{R}^n$ is defined as $\sgn(x) = [\sgn(x_1) \ldots \sgn(x_n)]$. 
$\R^+$ denotes non-negative reals. Kronecker product is denoted by $\otimes$. The Euclidean norm of a vector $x$ is denoted by $\norm{x}$ and the corresponding induced matrix Euclidean norm by $\|A\|$ for a matrix $A$. A diagonal matrix with elements $d_1,d_2,...,d_n$ on the diagonal is represented by $\diag(d_1,...,d_n)$. The $n\times n$ identity matrix and zero matrix are denoted by $I_n$ and $0_n$ respectively; a $n$-dimensional vector of ones is denoted $\one_n$. For a matrix $A$, the maximum and minimum eigenvalues are respectively denoted by $\lambda_{max}(A)$ and $\lambda_{min}(A)$. For a symmetric matrix $\Gamma$, the notation $\Gamma > 0$ ($\Gamma < 0$) is used to denote a positive-definite (negative-definite) matrix. For a matrix signal, $A(\cdot):\R^+ \rightarrow \R^{p_1 \times p_2} $, we define $\|A\|_\infty \triangleq \sup_{t \geq 0} \|A(t)\|$, the signal infinity norm. Time and initial condition arguments for all state variables (variables with dynamics) are uniformly omitted for notational simplicity. Similarly, function arguments for the control variables are suppressed, and they are made clear through explicit expressions proposed later in the manuscript.

\subsection{Graph Theory}

Consider a network of $n$ agents interacting with each other over a time-varying graph. 
% Let this network be modeled as  \textit{undirected} graph. 
We define the interaction graph as a function of time ($t \ge 0$) through the tuple, $\mathcal{G}(t) \triangleq (\mathcal{V},\mathcal{E}(t))$, where $\mathcal{V} \triangleq {1,...,n}$ is a node set and $\mathcal{E}(t) \subseteq \mathcal{V} \times \mathcal{V} $ is an edge set signifying interaction between nodes \cite{weiren} at time instant `$t$'. 
% An edge $(i,j)$ in the edge set of a directed graph signifies that agent $j$ can obtain information from agent $i$ but not vice-versa. 
If an edge $(i,j) \in \mathcal{E}(t)$, then node $i$ is called a \textit{neighbor} of node $j$ with $j$ being the \emph{head} node and $i$ being the \emph{tail} node indicating information flow from $i \rightarrow j$. The set of neighbors of a node $i$ at time $t$, is denoted by $\mathcal{N}_i(t)$. In an undirected graph, $(j,i) \in \mathcal{E}(t) \Leftrightarrow (i,j) \in \mathcal{E}(t)$ for all $i,j \in \mathcal{V}$. An undirected graph $\Gr(t)$ is instantaneously \emph{connected} if there is an undirected path between every pair of distinct nodes. The \textit{adjacency} matrix, $\R^{n \times n} \ni \Adj(t) = [a_{ij}(t)]$, is defined such that $a_{ij}(t) > 0$ if $(j,i) \in \mathcal{E}(t)$ and $a_{ij}(t) = 0$ if $(j,i) \notin \mathcal{E}(t)$. We assume no self edges are present and hence, $a_{ii}(t) = 0$ for all $t$. For an undirected graph, $\Adj$ is symmetric. The \textit{degree} matrix of the graph $\Gr$ is, $\Deg(t) \triangleq \diag(\sum \limits_{j=1}^{n} a_{1j}(t),...,\sum \limits_{j=1}^{n} a_{nj}(t)) \in \R^{n \times n}$ and the \textit{Laplacian} matrix, $\Lap(t) \triangleq [l_{ij}(t)] \in \R^{n \times n}$ is computed as: 
\begin{align*}
    \Lap(t) &= \Deg(t) - \Adj(t) \\
    l_{ii}(t) &= \sum\limits_{j=1,j \neq i}^n a_{ij}(t), \quad \quad l_{ij}(t) = -a_{ij}(t), i\neq j
\end{align*}
As evident from above, $\Lap(t)$ is symmetric for undirected graphs. Further, $\Lap(t)$ has both row and column sums zero indicating that $0$ is an eigenvalue with a corresponding eigenvector being $\one_n$ (vector of ones), i.e. $\Lap(t) \one_n = \one_n^\top \Lap(t) = 0$. Another symmetric matrix of interest is the \emph{signless} Laplacian defined as $\Lapu(t) \triangleq \Deg(t) + \Adj(t)\triangleq[Q_{ij}(t)] \in \mathbb{R}^{n \times n}$ where,
\begin{align*}
    Q_{ii}(t) &= \sum\limits_{j=1,j \neq i}^n a_{ij}(t), \quad \quad Q_{ij}(t) = a_{ij}(t), i\neq j.
\end{align*}
For an undirected graph $\Gr(t)$, both $\Lap(t)$ and $\Lapu(t)$ are positive semi-definite matrices. A \emph{union graph} denoted $\cup_{\tau\in[t,t+T]} \Gr(\tau)$ is the graph formed by Adjacency matrix elements, $\bar{a}_{ij}(t) \triangleq \int_t^{t+T} a_{ij}(\tau) \dif \tau$. The union graph, as defined, is the graph obtained by collecting all the edges in the sub-graphs appearing over a time-interval $[t,t+T]$.

% \hlt{$\mathcal{L}x$ is a column stack vector of $\sum \limits_{j=1}^n a_{ij}(x_i - x_j)$, where $x = [x_1,...,x_n]^T \in \mathbb{R}^n$.}

\begin{defn}\label{def:inc}
\cite{AST17} For a time-varying graph $\Gr(t)$ with adjacency matrix elements $a_{ij}(t)$, the \emph{weighted incidence matrix} $\mathcal{H}:\R^+ \rightarrow \R^{n \times \frac{n(n-1)}{2}}$ is defined as,
\begin{align*}
    \mathcal{H}(t) \triangleq h_{ij}(t) \triangleq \begin{cases}
        \sqrt{a_{ij}(t)}, &  \text{if }e_j = (i,j) \\
        -\sqrt{a_{ij}(t)}, & \text{if }e_j = (j,i) \\
        0, &\quad \textrm{otherwise}
    \end{cases}
\end{align*}`
\end{defn}
\begin{rem}
    In the aforementioned definition, for undirected graphs, it is standard practice to choose an arbitrary orientation (information flow direction). This has no effect on the graph Laplacian and can always be computed as, $\Lap(t) = \mathcal{H}(t)\mathcal{H}^T(t)$.
\end{rem}
\begin{defn}
\cite{ADH-98} At any given time `$t$', an undirected graph $\Gr(t)$ is called \emph{bipartite} if there exists a disjoint partition of the node set denoted as $\mathcal{V} = \mathcal{V}_+(t) \cup \mathcal{V}_-(t)$ such that all edges in $\Gr(t)$ are between the node sets, and there are no edges within the node set. Mathematically, for all $(i,j) \in \mathcal{E}(t)$, $i \in \mathcal{V}_k(t) \implies j \in \mathcal{V}\setminus \mathcal{V}_k(t)$ for $k \in \{+,-\}$. A graph is called \emph{jointly (non-)bipartite} over $[t,t+T]$ if the corresponding union graph $\cup_{\tau\in[t,t+T]} \Gr(\tau)$ is (non-)bipartite.
\begin{rem}
The above definition implies that the graph need not necessarily be  (non-)bipartite for all time instants between $[t,t+T]$, but the graph obtained by collecting all the edges in the sub-graphs appearing over the time interval is (non-)bipartite. 
\end{rem}

\end{defn}
\begin{defn}\label{def:dtcon}
\cite{AST17} The time-varying graph $\mathcal{G}(t)$ is termed
jointly $(\delta, T)$-connected if there are two real numbers $\delta > 0$
and $T > 0$ such that the edges $(j,i)$ satisfying,
\begin{align*}
    &\int_t^{t+T} a_{ij}(s) \dif s  \ge \delta, \quad i,j \in \mathcal{V}
\end{align*}
form a connected graph over $\mathcal{V}$ for all $t \geqslant 0$.
\end{defn}
\begin{defn}\label{def:PE}
(Persistence of Excitation) A locally integrable function $\phi:\R^+ \rightarrow \R^{n\times m}$ is said to be persistently exciting if there exist positive constants $\mu_1, \mu_2$, and $T$ such that,
\begin{align*}
    \mu_1 I_n \leqslant \int_t^{t+T} \phi(\tau) \phi^\top(\tau) \dif \tau \leqslant \mu_2 I_n, \quad \forall t\ge 0
\end{align*}
\end{defn}
\begin{defn}\label{def:IE}
\cite{sbasu} (Initial Excitation) A locally integrable function $\phi:\R^+ \rightarrow \R^{p\times q}$ is said to be initially exciting if there exist constants $\bar{T}$, $\eta$ $> 0$ such that,
\begin{align*}
    \int_{t_0}^{t_0+\bar{T}} \phi^\top(\tau) \phi(\tau) \dif \tau \geqslant \eta I_q, \quad \text{some } t_0 \geq 0
\end{align*}
\end{defn}
The extension of persistence and initial excitation conditions to multi-agent systems are termed collective persistence of excitation (C-PE) and collective initial excitation (C-IE) respectively \cite{garg2020distributed}. These are defined below.
\begin{defn}\label{def:C-PE}
A set of bounded, locally integrable signals $\phi_i:\R^+ \rightarrow \R^{r\times s}$, $\forall i=\{1,...,n\}$, are C-PE, if there exist constants $T>0$ and $\gamma>0$ such that,
\begin{equation*}
    \int_{t}^{t+T}\sum_{i=1}^{n}\phi_i(\tau)\phi_i(\tau)^{\top}d\tau \geq \gamma I_r, \quad \forall  t \geq t_0 \geq 0
\end{equation*}
\end{defn}
\begin{defn}\label{def:C-IE}
A set of bounded, locally integrable signals $\phi_i:\R^+ \rightarrow \R^{u\times v}$, $\forall i=\{1,...,n\}$, are C-IE, if there exist constants $\bar{T}>0$ and $\gamma>0$ such that,
\begin{equation*}
    \int_{t_0}^{t_0+\bar{T}}\sum_{i=1}^{n}\phi_i^{\top}(\tau)\phi_i(\tau)d\tau \geq \gamma I_v, \quad \text{some } t_0 \geq 0
\end{equation*}
\end{defn}
% \subsection{Model Description}

% The dynamics for agents in $\mathcal{G}$ is given by a subclass of Euler-Lagrangian (EL) equation,
% \begin{align}
%     \label{EL}
%     M_i(q)_i\ddot{q}_i + C_i(q_i,\dot{q}_i)\dot{q}_i + F_i(\dot{q}_i)\dot{q}_i = \tau
% \end{align}
% where $q \, \in \, \mathbb{R}^m$ is the vector of generalised coordinates, $M_i(q_i) \, \in \, \mathbb{R}^{m,m}$ is the inertia matrix, $C_i(q_i,\dot{q}_i)\dot{q}_i \, \in \, \mathcal{R}^{m}$ is the vector of Coriolis and centrifugal force, $F_i(\dot{q}_i)\dot{q}_i \, \in \, \mathcal{R}^{m}$ is the vector of frictional force and $\tau \, \in \, \mathcal{R}^{m}$ is the vector of control force. \eqref{EL} is a subclass of EL equation since the gravitational force term, $g_i(q_i)$, is not included in it. 
% We assume that the system described by \eqref{EL} have following properties,
% \begin{enumerate}
%     \item The matrix $M_i(q)$ symmetric positive-definite. 
%     \item There exist positive constants $k_{m_i} and k_{c_i} $ such that $M_i(q_i) - k_{m_i} \vb{I}_m \leq 0$ and $\norm{C_i(q_i,\dot{q}_i)\dot{q}_i } \leq k_{c_i}$.
%     \item $\dot{M}_i(q) - 2C_i(q_i,\dot{q}_i)$ is skew symmetric.
% \end{enumerate}
The following assumption is intrinsic to the subsequent results.

\begin{assum}
 \label{ass_g}
 
 The network graph, $\Gr(t)$, is undirected and jointly $(\delta, T)$-connected for some $\delta, \, T >0$. We assume the same graph $\Gr(t)$ for both relative measurements as well as information exchange. Further, we assume existence of an $a_M >0$ such that $a_{ij}(t) \leqslant a_M$ for all $i,j \in \{1,2,\ldots,n\}$ and for all $t \geqslant 0$.
\end{assum}

\subsection{Fundamental Results}

We state a few results from graph theory and consensus analysis to be used subsequently.
\begin{prop}
\cite{ADH-98} A graph $\Gr$ is bipartite if and only if $\Gr$ has no cycle of odd length.
\end{prop}
\begin{prop}\label{thm:bip}
\cite{CRS-07} The smallest eigenvalue of the signless
Laplacian matrix $\Lapu = \Deg + \Adj$ of an undirected and connected graph is equal to zero if and only if the graph is bipartite. In case the graph is bipartite, zero is a simple eigenvalue
\end{prop}
The following is a re-wording of \cite[Theorem 3.4]{RSML-18}.
\begin{prop} \label{thm:peV}
Consider the time-varying dynamics,
\begin{align}\label{eq:pesys}
    \dot{x} = -\sigma N(t)N^\top (t) x, \quad x(0) = x_0
\end{align}
with $N: \R^+ \rightarrow \R^{k \times p}$ being a piecewise continuous matrix function. If $N(\cdot)$ is persistently exciting (\Cref{def:PE}), then the above dynamics admits a Lyapunov function
\begin{align*}
    & V(t,x) = \frac{1}{2}x^\top [\pi I_k + S(t)]x
\end{align*}
where,
\begin{align*}
    & S(t) = 2\delta_T I_k - \frac{2}{T}\int_t^{t+T} \int_t^{r} N(\tau)N^\top(\tau) \dif \tau \dif r
\end{align*}
and the positive constants $\pi, \delta_T$ are defined as,
\begin{align*}
    & \delta_T \triangleq T | N(\cdot)N^\top(\cdot) | _\infty, \\
    & \pi \triangleq 1 + \frac{2\sigma^2\delta_T^3}{\mu_1}.
\end{align*}
Further, the states of the dynamical system~\eqref{eq:pesys} are uniformly exponentially stable at the origin.
\end{prop}
The following result was established as part of the proof in \cite[section IIA]{AST17}.

\begin{prop} \label{thm:con-pe}
The following hold for an undirected graph $\Gr(t)$ over $n$- nodes, with Laplacian $\Lap(t)$ and weighted incidence matrix $\mathcal{H}(t)$ (\Cref{def:inc}).
\begin{itemize}
    \item  $\Big(\Lap(t) + \frac{\one_n\one_n^\top}{n} \Big) = \Big(\mathcal{H}(t) + \frac{\one_n h^\top(t)}{\sqrt{n}}\Big)\Big(\mathcal{H}(t) + \frac{\one_n h^\top(t)}{\sqrt{n}}\Big)^\top $, where $h(t)$ is unit vector in the kernel of $\mathcal{H}(t)$.
    \item The graph $\Gr(t)$ is jointly $(\delta,T)$-connected (\Cref{def:dtcon}) for some $\delta ,T > 0$ if and only if $\Big(\mathcal{H}(t) + \frac{\one_n h^\top(t)}{\sqrt{n}}\Big)$ is persistently exciting (\Cref{def:PE}).
\end{itemize}
\end{prop}

\section{Bias Estimated Consensus}

The objective of this article is to develop a distributed consensus algorithm for a network of double integrator systems in the presence of a constant unknown bias corrupting the relative measurements of position while ensuring estimation of all biases by each agent. The interaction between the agents is modeled by an undirected and jointly $(\delta, T)$-connected graph, $\Gr(t)$, with an associated Laplacian $\Lap(t)$. The input-output model for each agent representing a node in $\Gr(t)$ is expressed as the following double integrator equation,
\begin{align}
    \label{EL}
    &\ddot{q}_i=u_i, \nonumber\\
     &q_i(0) = q_{i0}; \, \dot{q}_{i}(0) = \dot{q}_{i0};\, i=1,2,\ldots, n \nonumber \\
  %& \dot{x}_{1_i}=x_{2_i},\nonumber\\
    %& \dot{x}_{2_i}=u_i, \quad i=1,2,\ldots, n \nonumber \\
     &y_i = [z_{ij},\; \dot{q}_i]^{\top}, \quad i=1,2,\ldots, n,\, j \in \mathcal{N}_i(t)
\end{align}
 where state $q_i \, \in \, \mathbb{R}^m$ is the vector of generalized coordinates (called `positions' in general with their derivatives being `velocities'), $u_i(\cdot) \in \R^{m}$ is a distributed, time-varying feedback, $y_i=[z_{ij},\; \dot{q}_{i}] \, \in \, \mathbb{R}^{2m}$ is the information state available to each agent with $\R^m \ni z_{ij} \triangleq (q_i - q_j + b_i)$ and $b_i \, \in \, \mathbb{R}^{m}$ is the \emph{constant, unknown} sensor bias. The estimate of sensor bias ($b_i$) for each agent $k$ will be denoted $\hat{b}_i^k$ in the sequel. Each agent $k$ has an estimate of all sensor biases ($\hat{b}^k = [\hat{b}_1^k,\hat{b}_2^k,\ldots,\hat{b}_n^k]^\top)$ and a dynamic update law is designed for the same.
% \eqref{EL} is a subclass of EL equation since the gravitational force term, $g_i(q_i)$, is not included in it. The gravity component is not considered in the dynamics since we are considering robot manipulators moving in the horizontal direction. 

The control objective in this article is to design a distributed feedback $u_i$ so that the closed-loop solutions of \eqref{EL} satisfy,
\begin{align}
    &\lim_{t \to \infty} (q_i-q_j) = 0, \quad \forall i,j \in \{1,2,\ldots, n\} \nonumber\\
    &\lim_{t \to \infty} \dot{q}_i = 0 \quad \forall i \in \{1,2,\ldots, n\} \nonumber\\
    &\lim_{t \to \infty} (b_i-\hat{b}_i^k) = 0 \quad \forall i,k \in \{1,2,\ldots, n\}
    \label{eq:obj}
    \tag{Bias Estimated Consensus}
\end{align}

The following assumption delineates the information available for the design of the feedback law $u_i$.
\begin{assum}
\label{ass_meas}
Agents can measure their own velocities ($\dot{q}_i$); neighbors measure relative velocities ($\dot{q}_i - \dot{q}_j$) and relative position corrupted by a constant unknown bias ($z_{ij} = q_i - q_j + b_i$). Further, neighbors exchange their measurement of relative positions ($z_{ji} = q_j - q_i + b_j$) and their estimate of biases ($\hat{b}^j$) with each other.
\end{assum}

\section{Control Law Design}

We prescribe a controller $u_i$ for \eqref{EL} to satisfy our control objective \eqref{eq:obj} where $b_i, \, i = 1,2,\ldots,n$ are assumed to be \emph{unknown, constant} measurement biases.
%Following assumptions are considered in addition to assumptions \ref{ass_g} to \ref{ass_meas},
% \begin{assum}
% \label{ass_traj}
%     The trajectory of the leader $q_0 \in \mathcal{C}^2$ and $q_0,\dot{q}_0,
%     \ddot{q}_0 \in \mathcal{L}_{\infty}$. Furthermore, it is assumed that $\dot{q}_0 \; \mathtext{and} \; \ddot{q}_0$ are known to the followers.
% \end{assum}
%     \begin{rem}
%         Assumption \ref{ass_traj} is reasonable for the scenarios when the target is fixed (zero velocity) or when the leader is artificial (knowledge of $\ddot{q}_0$). The latter case is then equivalent to all followers being connected to the artificial leader.
%     \end{rem}
%\begin{assum}
%\label{ass:md}
%The matrix functions $M_i(q_i)$, $C_i(q_i,\dot{q}_i)$, $g_i(q_i)$, $F_i(\dot{q}_i)$ are assumed to be known to the $i^{th}$ agent. \edt{Further, there exist constants $\alpha_m, \alpha_M > 0$ such that, $ \alpha_m I_m \leqslant M_i(q_i) \leqslant \alpha_M I_m$ for all $q_i$ and $i \in \{1,2,\ldots,n\}$.}
%\end{assum}
% \begin{rem}
%     Assumption \ref{ass:md} requires that self information be available without bias along with typical system parameters such as mass, inertia, friction coefficients etc.  
% \end{rem}
\noindent
%We can re-write \eqref{EL}  with the position and velocity vectors to be $q_i$ and $\dot{q}_i$ respectively as,
    %\begin{equation} \label{qddot}
    %\ddot{q}_i=u_i
    %\end{equation}
%\begin{align}
%\label{qddot}
    %\ddot{q}_i = M^{-1}_i(q_i)\big(\tau_i-C_i(q_i,\dot{q}_i)\dot{q}_i-F_i(\dot{q}_i)\dot{q}_i - g_i(q_i)\big)
%\end{align}

\noindent
In this article, we consider the following distributed control algorithm,
\begin{align*}
     u_i = k(t) \Big(-\dot{q}_i 
     -\frac{1}{2} \sum_{j \in \mathcal{N}_i}a_{ij}(t)[z_{ij}+z_{ji}]\Big) + w_i
\end{align*}
%\begin{align*}
 %   & u_i = k(t) \Big(-\dot{q}_i - \sum_{j \in \mathcal{N}_i} a_{ij}(t)(q_i-q_j+b_i)\\
  %  & 
   %  + \frac{1}{2} \sum_{j \in \mathcal{N}_i}a_{ij}(t)[(q_i-q_j+b_i)-(q_j-q_i+b_j)]\Big) + w_i 
     %+ C_i(q_i,\dot{q}_i)\dot{q}_i +F_i(\dot{q}_i)\dot{q}_i + g_i(q_i)%\\
    % simplifying the above equation
%\end{align*}
%\begin{align*}
 %    &= k(t)\Big(-\dot{q}_i + \cancel{\sum_{j \in \mathcal{N}_i}a_{ij}(q_i-q_j)} -\cancel{\sum_{j \in \mathcal{N}_i}a_{ij}(t)(q_i-q_j)} \\
  %   &  + 
   % \frac{1}{2} \sum_{j \in \mathcal{N}_i}a_{ij}(t)(b_i-b_j) - \sum_{j \in \mathcal{N}_i}a_{ij}(t)\,b_i \Big) 
   %  + w_i %+ C_i(q_i,\dot{q}_i)\dot{q}_i +F_i(\dot{q}_i)\dot{q}_i + g_i(q_i) 
    %\end{align*}
    where, $k: \R^+ \rightarrow \R^+ $ is a positive valued uniformly bounded function (there exists $k_M > 0$ such that $k(t) \leqslant k_M$ for all $t \geqslant 0$) to be prescribed later, $q = [q_1, \ldots, q_n]^\top$ and $w_i(\cdot) \in \R^m$ is an auxiliary control term. The implementation of the term $z_{ji}$ %$(q_j-q_i+b_j)$ 
    in the control law requires that all neighbors' relative position measurements (corrupted by bias) be communicated to each agent. This is guaranteed by \Cref{ass_meas}. The individual control expressions above can now be collected to specify the feedback for the entire double integrator network as follows:
    \begin{align}
    \label{tau}
    &u  = -k(t)\dot{q}  + \frac{k(t)}{2}[\Lap(t) \otimes I_m ]b - k(t)[\Deg(t) \otimes I_m]b + w
\end{align}
where, $u,\; b \; \mathtext{and} \; w$ are the column stacked vectors of $[u_1, \ldots, u_n]^\top$, $[b_1, \ldots, b_n]^\top$ and $[w_1, \ldots, w_n]^\top $ respectively. %$C(q,\dot{q}) = \diag(C_1(q_1,\dot{q}_1), \ldots, C_n(q_n,\dot{q}_n))$, and $F(\dot{q}) = \diag(F_1(\dot{q}_1), \ldots, F_n(\dot{q}_n))$, $g(q) = [g_1(q_1), \ldots, g_n(q_n)]^T$. 
Let $\bar{\Lap}(t)\triangleq \Lap(t) \otimes I_m \; \mathtext{and} \; \bar{\Deg}(t) := \Deg(t) \otimes I_m $, then \eqref{tau} can be simplified as,
\begin{align}
    \label{tau_final}
    u  &= -k(t)\bar{\Deg}(t)b - k(t)\dot{q}+\frac{k(t)}{2}\bar{\Lap}(t)b + w%+C(q,\dot{q})\dot{q}+F(\dot{q})\dot{q} + g(q)
\end{align}
We also obtain the network dynamics from \eqref{EL} as,
\begin{align}
    \label{qddot_aug}
    \ddot{q} &= u.
\end{align}
Substituting the control law \eqref{tau_final} in \eqref{qddot_aug}, we obtain the following closed-loop network dynamics,
\begin{align}
% \label{qddot_modified}
%     \ddot{q} &= M^{-1}\bigg(-k(t)\bar{D}b - k(t)\bar{L}\dot{e}+\frac{k(t)}{2}\bar{L}b+v+\cancel{C\dot{q}}+\cancel{F\dot{q}} +\cancel{g} -\cancel{C\dot{q}}-\cancel{F\dot{q}} -\cancel{g}\bigg) \\
    \label{new_dyn}
    \ddot{q} &+ k(t)\dot{q} + k(t)\bigg(\bar{\Deg}(t)-\frac{1}{2}\bar{\Lap}(t)\bigg)b = w.
\end{align}
 It is worth noting that $\bar{\Deg}(t)-\bar{\Lap}(t)/2 =  \Lapu(t)/2 \otimes I_m \triangleq \bar{\Lapu}(t)/2$.  \eqref{new_dyn} can be written in a standard regressor-parameter form as,
\begin{align*}
    & \underbrace{\Big[
    \ddot{q}, \;\; k(t)\dot{q}, \;\; \frac{1}{2}k(t)\bar{\Lapu}(t)
    \Big]}_{Y\in \mathbb{R}^{mn \times (mn+2)}}\underbrace{\Big[
    1 \; 1 \; b \Big]^T}_{\theta \in \mathbb{R}^{mn+2}} = w
\end{align*}
with $Y:\R^m \times \R^m \times \R \rightarrow \R^{mn \times (mn+2)}$ denoting the regressor and $\theta \in \R^{mn+2}$ being the constant, unknown parameter. 
%\hlt{\begin{align}
%   \label{regressor}
%   & \underbrace{\Big[
%  \ddot{q}, \;\; k(t)\dot{q}, \;\;\frac{1}{2}k(t)\bar{\Lapu}(t)
 %\end{align}}
%\hlt{with $Y:\R^m \times \R^m \times \R \rightarrow \R^{mn \times (mn+2)}$ which denotes the regressor . For each agent,  $Y_{i} \in \mathbb{R}^{m \times (mn+2)}$ corresponds to each $m$ rows of $Y$.}\\
%\hlt{The signless Laplacian, $\mathcal{Q}(t)=[Q_{ij}(t)] \in \mathbb{R}^{n \times n}$ is computed as:}
%\hlt{\begin{align*}
 % \mathcal{Q}&=\mathcal{D}+\mathcal{A}  \\
 % Q_{ii}(t)&=\sum_{j=1, j \neq i}^na_{ij}(t), \quad Q_{ij}(t)=a_{ij}(t),i\neq j
%\end{align*}}
We can also write, corresponding to each agent:
\begin{equation}\label{Yi}
\small{  Y_i=  \Big[
    \ddot{q}_i, \;\; k(t)\dot{q}_i, \;\; \frac{1}{2}k(t)[Q_{i1}(t)I_m,\;Q_{i2}(t)I_m,...,\;Q_{in}(t)I_m]
    \Big]}.
\end{equation}
$Y_{i}$ and $w_{i}$ are available for each agent, i.e., $Y_{i} \in \mathbb{R}^{m \times (mn+2)}$ corresponds to each $m$ rows of $Y$, $w_{i} \in \mathbb{R}^{m}$ corresponds to each $m$ rows of $w$. Additionally, we have corresponding to each agent, $Y_{i}\theta=w_{i}$. Each agent has an adaptive estimate of the unknown parameter vector  $\theta$ for all $i=1,..,n$ (which is an over-parametrization of $b$) denoted $\hat{\theta}^i \in \mathbb{R}^{mn+2}$. $\hat{\theta}^i =[p^i \;l^i\;\hat{b}^i]^{\top}$ where $\hat{b}^i=[\hat{b}^i_1,\hat{b}^i_2,...,\hat{b}^i_n]^{\top}$. %The estimate of bias $b_1,..,b_n$ computed by agent $i$ is known to agent $i$.
$p^i$, $l^i$ are the estimates of the constants in $\theta$ by agent $i$. We define the agent parameter error as, $\tilde{\theta}^i \triangleq [1-p^i\;1-l^i\; \tilde{b}^i]^{\top}$, where $\tilde{b}^i \triangleq [b_1-\hat{b}_1^i,...,b_n-\hat{b}_n^i]^{\top}=[\tilde{b}_1^i,...,\tilde{b}_n^i]^{\top}$.
We now define,
\begin{align}
\label{s_i}
    s_i \triangleq \dot{q}_i + \lambda \bigg(q_i + 
     \frac{\tilde{b}_i^i}{2}\bigg) \gap \lambda > 0, \,i =1,2,\ldots n
\end{align}
where $\tilde{b}_i^i \triangleq b_i-
\hat{b}_i^i$. Further we assign, $s \triangleq [s_1,\,\ldots,\,s_n]^\top$, $\tilde{b}_{new} \triangleq [\tilde{b}_1^1,\,\ldots,\,\tilde{b}_n^n]^{\top}$ and obtain $s = \dot{q} + \lambda (q + \tilde{b}_{new}/2)$.\\\\
\textbf{Note} - $\tilde{b}_i^i$ has been used in \cref{s_i}, and not  $\tilde{b}_i^k$  as $\tilde{b}_i^k=b_i-\hat{b}_i^k$ will require bias estimate information of $b_i$ computed by agent $k \not\in \mathcal{N}_i$, which may not be available with agent $i$.\\
Taking the derivative of $s$ and substituting from \eqref{new_dyn},

\begin{align}
% \label{stilde_dot}
    &\dot{s} =  \ddot{q} +\lambda\bigg(\dot{q} +\frac{\dot{\tilde{b}}_{new}}{2}\bigg)\nonumber\\
    &= -k(t)\dot{q} - \frac{1}{2}k(t)\bar{\Lapu}(t)b + w+\lambda\bigg(\dot{q}+\frac{\dot{\tilde{b}}_{new}}{2}\bigg) \nonumber\\
% \end{align}
% \begin{align}
\label{stilde_dot}
    % \dot{\tilde{s}}
    &=\underbrace{\begin{bmatrix}
    0_{mn\times 1}&0_{mn\times 1}&- \frac{1}{2}k(t)\bar{\Lapu}(t)
    \end{bmatrix}}_{Z \in \mathbb{R}^{mn \times (mn+2)}}\begin{bmatrix}1&1&b\end{bmatrix}^\top\nonumber \\
    & -k(t)\dot{q}+w+\lambda\bigg(\dot{q}+\frac{\dot{\tilde{b}}_{new}}{2}\bigg).
\end{align}
The corresponding dynamics for each agent $s_i\in \mathbb{R}^m$, is given by
\begin{equation*}\label{si}
\dot{s}_i=Z_i\theta-k(t)\dot{q}_i+w_i+\lambda\Big(\dot{q}_i+\frac{\dot{\tilde{b}}_i^i}{2}\Big)
\end{equation*}
where $Z_i \in \mathbb{R}^{m \times(mn+2)}$ is defined similar to $Y_i$.
We now define the second part of the control, $w$
%{\edt{
%\begin{align}
%\label{u}
%    w & \triangleq k(t)\dot{q} -Z\hat{\theta}-\sigma\bar{\Lap}(t)s -\lambda \dot{q} -\frac{\lambda\dot{\tilde{b}}_{new}}{2},\, \sigma > 0
%\end{align}}}
at each agent node as,
\begin{align}\label{w}
    w_i=k(t)\dot{q}_i-Z_i\hat{\theta}^i-\lambda(\dot{q}_i+\frac{\dot{\tilde{b}}^i_i}{2})-\sum_{j\in \mathcal{N}_i}a_{ij}(t)(s_i-s_j),
\end{align}
which can be written in an implementable form as,
\begin{align} \label{wi}
    & w_i = k(t)\dot{q}_i-\lambda\dot{q}_i -\frac{\lambda\dot{\tilde{b}}_i^i}{2}  +k(t)\sum\limits_{j \in \mathcal{N}_i} a_{ij}(t)\hat{b}_i^i  \nonumber \\
    & - \sigma\sum\limits_{j \in \mathcal{N}_i} a_{ij}(t)(s_i-s_j) -\frac{k(t)}{2}\sum\limits_{j \in \mathcal{N}_i} a_{ij}(t)(\hat{b}_i^i - \hat{b}_j^i) ,\, \sigma > 0.
\end{align}
The above yields for the entire network the following,
\begin{align} \label{wfull}
    w&=k(t)\dot{q}-\lambda(\dot{q}+\frac{\dot{\tilde{b}}_{new}}{2})-\sigma\bar{\Lap}(t)s-Z_{new}\hat{\theta}
\end{align}
where $\hat{\theta}:=[\hat{\theta}^1,..,\hat{\theta}^n]^{\top}$ and $Z_{new}(\dot{q},t)\in \mathbb{R}^{mn \times n(mn+2)}$ is defined as
\begin{align*}Z_{new}(\dot{q},t)\triangleq \diag(Z_1(\dot{q}_1,t),Z_2(\dot{q}_2,t),..,Z_n(\dot{q}_n,t)).\end{align*}

Since the bias, $b$, is constant we have, $\dot{\tilde{b}}^i_i = -\dot{\hat{b}}_i^i$ and hence is implementable in \eqref{wi}. Further, though $s_i$ is not implementable (due to the $(q_i + \tilde{b}_i^i/2)$ term in $s_i$), $(s_i-s_j) = \dot{q}_i - \dot{q}_j + \frac{1}{2}\lambda(z_{ij} - z_{ji} - (\hat{b}_i^i-\hat{b}_j^j) )$ is, and that is what appears in the control law~\eqref{wi}. Further, the implementation of the term $(\hat{b}_i^i-\hat{b}_j^j)$ requires neighbors to exchange their bias estimates (Assumption \ref{ass_meas}).

The control law $u_i$ after substituting for $w_i$ from \eqref{wi} is given by:-
    \begin{equation*}
    \begin{aligned}
    u_i&=\frac{k(t)}{2}\sum_{j \in \mathcal{N}_i} a_{ij}(t)(\tilde{b}_i^i-\tilde{b}_j^i)-\lambda \dot{q}_i-\frac{\lambda \dot{\tilde{b}}_i^i}{2}\\
    & -\sigma \sum _{j \in \mathcal{N}_i} a_{ij}(t)(s_i-s_j)
     -k(t)\sum_{j \in \mathcal{N}_i} a_{ij}(t)(\tilde{b}_i^i)
\end{aligned}
\end{equation*}

%where $\hat{\theta}^i$ is the adaptive estimate of $\theta$ which is an over-parametrization of $b$ for each agent $i$ as defined in \eqref{define_theta}. 
Substituting \eqref{w} in \eqref{stilde_dot},
\begin{align}\label{eq:clsys}
    \dot{s} &= Z_{new}\tilde{\theta}-\sigma\bar{\Lap}(t)s
\end{align}
where $\tilde{\theta} := [\tilde{\theta}^1,...,\tilde{\theta}^n]^{\top}$, $\tilde{\theta}^i=\theta-\hat{\theta}^i$.
%With the above defined notation, we now have,

\subsection{Bias Estimation}
The matrix $Y_i(\dot{q}_i ,\ddot{q}_i,t)$ in \eqref{Yi} is dependent on the acceleration term $\ddot{q}_i$ and so cannot be used in our adaptation law for bias estimation. In order to facilitate relaxation of the persistence of excitation condition, a filter is designed for each agent as proposed in \cite{garg2020distributed},

\begin{equation}
\begin{aligned}
\label{yfdot}
    \dot{Y}_{F_i} &= -\beta Y_{F_i}+Y_i(\dot{q_i} ,\ddot{q_i},t) \gap Y_{F_i}(0) = 0\\
    \dot{w}_{F_i} &= -\beta w_{F_i}+ w_i \gap w_{F_i}(0) = 0
\end{aligned}
\end{equation}
where $\beta > 0$ is the scalar filter gain, $Y_{F_i} \in \mathbb{R}^{m\times (mn+2)}$ and $w_{F_i} \in \mathbb{R}^{m}$.  Solving the above equations explicitly we obtain,
\begin{align}
\label{yf}
    Y_{F_i}(t) &= e^{-\beta t}\int_{0}^{t}e^{\beta r}Y_i(\dot{q_i} ,\ddot{q_i},r)\dif r\\
    \label{uf}
    w_{F_i}(t) &= e^{-\beta t}\int_{0}^{t}e^{\beta r}w_i \dif r.
\end{align}
Utilizing the relation $Y_i(\dot{q}_i ,\ddot{q}_i,t)\theta=w_i$ we get $Y_{F_i}\theta = w_{F_i}$ from \eqref{yf} and \eqref{uf}. $Y_{F_i}$ in \eqref{yfdot} cannot be solved explicitly as $Y_i(\dot{q}_i ,\ddot{q}_i,t)$ is not measured. Therefore, we split $Y_i(\dot{q}_i ,\ddot{q}_i,t)$ into measured and non-measured parts as,
 \begin{align*}
     Y_i(\dot{q}_i ,\ddot{q}_i,t) &= Y_{1_i}(\ddot{q}_i) + Y_{2_i}(\dot{q}_i , t)
 \end{align*}
where
\begin{align*}
    & Y_{1_i}(\ddot{q}_i) = \Big[
    \ddot{q}_i, \;\; 0_{m\times1}, \;\; 0_{m\times mn}
   \Big]  \\
    & Y_{2_i}(\dot{q}_i , t) = \\
    & \Big[
    0_{m\times1}, \;\; k(t)\dot{q}_i, \;\; \frac{1}{2}k(t)[Q_{i1}(t)I_m,\;Q_{i2}(t)I_m,...,\;Q_{in}(t)I_m]
    \Big]
\end{align*}
This implies that $Y_{F_i} = Y_{F_{1_i}}+Y_{F_{2_i}}$, where,
\begin{equation}
\begin{aligned}
    \dot{Y}_{F_{1_i}} &= -\beta Y_{F_{1_i}} + Y_{1_i} \gap Y_{F_{1_i}}(0) = 0\\
    \label{yf2_dot}
    \dot{Y}_{F_{2_i}} &= -\beta Y_{F_{2_i}} + Y_{2_i} \gap Y_{F{2_i}}(0) = 0
\end{aligned}
\end{equation}
Since $Y_{2_i}(\dot{q}_i , t)$ is known, $Y_{F_{2_i}}$ can be solved online using \eqref{yf2_dot} by employing a numerical integration scheme. We solve $Y_{F_{1_i}}$ analytically as follows,
\begin{align*}
    Y_{F_{1_i}}(t) &= e^{-\beta t}\int_0^te^{\beta r}{Y_{1_i}}(\ddot{q}(r))dr \nonumber\\
    &= \begin{bmatrix}
    e^{-\beta t}\int_0^te^{\beta r}\ddot{q}_i(r)dr & 0 & 0
    \end{bmatrix}
\end{align*}
The elements of $Y_{F_{1_i}}$ can be evaluated using integration by parts as follows,
\begin{align*}
   Y_{F_{1_i}}(t) &=  \scriptstyle{\begin{bmatrix}
    e^{-\beta t}[e^{\beta t}\dot{q}_i(t)-\dot{q}_i(0)]-e^{-\beta t}\int_0^t \beta e^{\beta r}\dot{q}_i(r)dr & 0 & 0\\
    \end{bmatrix}} \nonumber\\
     &= \begin{bmatrix}
    \dot{q}_i(t)-e^{-\beta t}\dot{q}_i(0)-h_i(t) & 0 & 0
    \end{bmatrix}
\end{align*}
 $\forall \; i = 1,\cdots n$ and 
\begin{align*}
    \dot{h}_i &= \beta \dot{q}_i-\beta h_i \gap h_i(0)=0
\end{align*}
%\comment{}{what are these j in the superscripts?}
%\comment{}{We have $\ddot{q}_i\in \mathbb{R}^m$, so $\ddot{q}_i=\begin{pmatrix} \ddot{q}_i^1\\ \vdots \\\ddot{q}_i^m \end{pmatrix}$. Each element of $\ddot{q}_i$ is represented by $\ddot{q}_i^j$, where $j$ varies from $1$ to $m$ and $i$ varies from $1$ to $n$. Here, either we can replace $j$ with some other letter (as $j$ represents neighbouring agents) or we can just remove the superscripts as $\ddot{q}_i$ implies we are taking the whole column vector, and similarly for $h_i$ dynamics.}
$Y_{F_i}$, $w_{F_i}$ are filtered regressor and filtered control for each agent $i$ respectively. $Y_{F_i}$ can now be used in our adaptation law. Additionally, for bias estimation, we will make use of the double filtered regressor and control law introduced in \cite{sbasu}, \cite{garg2020distributed}
\begin{align}
    \label{proj_yif}
    \dot{Y}_{{IF}_i} &= Y_{F_i}^{\top}Y_{F_i} \gap Y_{{IF}_i}(0) = 0 \\
\label{proj_uif}
    \dot{w}_{{IF}_i} &= Y_{F_i}^Tw_{F_i}  \gap w_{{IF}_i}(0) = 0
\end{align}
where $Y_{{IF}_i}\in \mathbb{R}^{(mn+2) \times(mn+2)}$, $w_{{IF}_i} \in \mathbb{R}^{(mn+2)}$.

\begin{fact} \cite{sbasu} \label{fact_Yw}
Integrating \eqref{proj_yif} and \eqref{proj_uif} and using the relation $Y_{F_{i}}\theta = w_{F_{i}}$ it can be verified that,
\begin{align}
\label{prop1}
    Y_{{IF}_i}\theta = w_{{IF}_i} \gap \forall t \geqslant 0
\end{align}
\end{fact}
\begin{fact}\cite{sbasu}\label{fact_nondecreasing}
The solution $Y_{{IF}_i}(t)$ of \eqref{proj_yif} is a non-negative and non-decreasing function of time.
% positive semi-definite function of time $i.e.$ $Y_{IF}(q(q_0,t),\dot{q}(q_0,t),t) \geq 0 $ $\forall t \geq 0$, $\forall q_0$
\end{fact}
% \begin{property}\cite{sbasu}
% $Y_{IF}(q(q_0,t),\dot{q}(q_0,t),t)$ is a non-decreasing function of time $i.e.$ $Y_{IF}(q(q_0,t),\dot{q}(q_0,t),t_2) - Y_{IF}(q(q_0,t),\dot{q}(q_0,t),t_1) \geq 0 \; \forall \;t_2 \geq t_1$, $\forall q_0$
% \end{property}
\noindent
% For proof see \cite{sbasu}. 
The adaptive control law for bias estimation is now chosen as,
\begin{align}
\label{ac_law}
\dot{\hat{\theta}}^i &=  \mu_FY_{F_i}^\top(w_{F_i}-Y_{F_i}\hat{\theta}^i) + \mu_{IF}(w_{{IF}_i}-Y_{{IF}_i}\hat{\theta}^i)\nonumber\\
&+\sum_{j\in \mathcal{N}_i}a_{ij}(\hat{\theta}^i-\hat{\theta}^j)
%&+\sum_{j\in \mathcal{N}_i}a_{ij}(\hat{\theta}_i-\hat{\theta}_j) 
, \forall i=1,..,n
\end{align}
which using \Cref{fact_Yw} can be written as,
\begin{align}
    \dot{\tilde{\theta}} &= -\mu_F\phi_F\tilde{\theta}-\mu_{IF}\phi_{IF}\tilde{\theta}-\mathcal{L}\otimes I_{(mn+2)}\tilde{\theta}
    %-(\mathcal{L}\otimesI_{(mn+2)})\tilde{\theta}
    \label{ac_law2}\\
    \dot{\hat{b}}^i &= [\dot{\hat{\theta}}^{i}_{(3)} \, \, \dot{\hat{\theta}}^{i}_{(4)} \, \, \cdots \,\, \dot{\hat{\theta}}^{i}_{({mn+2})}]^T \nonumber
\end{align}
for constant $\mu_F,\;\mu_{IF} > 0$ and arbitrary initial conditions. $\dot{\hat{\theta}}^{i}_{(k)}$ denotes the $k^{th}$-element of $\dot{\hat{\theta}}^i$ and so on.  $\phi_F,\phi_{IF} \in \mathbb{R}^{n(mn+2) \times n(mn+2)} $ are block diagonal matrices and are defined as,
\begin{align*}
  \phi_F&\triangleq \diag(Y_{F_1}^{\top} Y_{F_1},...,Y_{F_n}^{\top} Y_{F_n})\\
  \phi_{IF}&  \triangleq \diag(\int_0^tY_{F_1}^{\top} Y_{F_1},..,\int_0^tY_{F_n}^{\top} Y_{F_n}).
\end{align*}
%\hlt{\begin{rem}
%    There is no consensus term in the bias estimation error dynamics \cref{ac_law} because of the inherent consensus being achieved due to $\bar{\mathcal{Q}}(t)$ present in $Y_{F_i}$'s.
%\end{rem}}
Further, we consider the following assumption and a corresponding proposition.
\begin{assum}\label{ass_cie}
The set of filtered regressors $Y_{F_i}$ are C-IE as per \Cref{def:C-IE}.
\end{assum}
%\begin{assum}
%\label{ass_ie}
%$Y_F$ is initially exciting (IE) (\Cref{def:IE}) during the interval $[0,\bar{T}]$, with the degree of excitation $\eta>0$.
%\end{assum}
%\hlt{\begin{assum}\label{ass_yfibound}
%$Y_{F_i}(t)\in\mathcal{L}_2 \cap \mathcal{L}_{\infty}$, $\dot{Y}_{F_i} \in \mathcal{L}_{\infty}$, and $w_{F_i}\in\mathcal{L}_2 \cap \mathcal{L}_{\infty}$ forall $i=1,..,n$
%\end{assum}}

\begin{rem}
    It is worth noting that the solution $Y_{F_i}$ in the assumption above depends on the initial conditions of the closed-loop state $\dot{q}_i(0)$. The collective initial excitation condition is therefore not necessarily uniform with respect to initial data.
\end{rem}
\begin{prop}\label{garglemma}
\cite{garg2020distributed} Provided \Cref{ass_cie} holds, the matrix $M(t)=\mathcal{L}\otimes I_{(mn+2)}+\mu_{IF}\phi_{IF}$ appearing in \eqref{ac_law2} is uniformly strictly positive definite over the time window $[\bar{T},\infty)$ i.e.,
\begin{equation*}
    \xi^{\top}M(t)\xi>0,\gap \forall t\geq \bar{T}
\end{equation*}
$\forall \xi \in \mathbb{R}^{n(mn+2)}.$
\end{prop}
We are now ready to state the primary result of this article.

\begin{thm} \label{thm:main}
Consider the multi-agent network with the agent dynamics given by \eqref{EL} interacting over an undirected graph $\Gr(t)$. If Assumptions \ref{ass_g}-\ref{ass_cie} hold, then the control law given by,
\begin{align*}
%  \tau  &= -k(t)\bar{\Deg}b - k(t)\bar{\Lap}\dot{q}+\frac{k(t)}{2}\bar{\Lap}b\nonumber \\
%     & +v(q,\dot{q},t)+C(q,\dot{q})\dot{q}+F(\dot{q})\dot{q} + g(q)
    u &=Z_{new}{\tilde{\theta}}-\sigma\bar{\mathcal{L}}(t)s-\lambda \dot{q}-\lambda \frac{\dot{\tilde{b}}_{new}}{2}
    \end{align*}
    %&= k(t)\Big(-\bar{\Lap}(t)\dot{q} - %\frac{1}{2}\bar{\Adj}(t)(z_{ij}+z_{ji}) \Big) \nonumber \\
   % &+C(q,\dot{q})\dot{q}+F(\dot{q})\dot{q} + g(q)+ M(q)w
%\end{align} 
%with $k: \R^m \times \R^+ \rightarrow \R $ being a positive valued function, $w$ as in \eqref{u},
with bias adaptation law \eqref{ac_law}, guarantees that $\lim \limits_{t \to \infty} (q_i-q_j) = 0$ (for all $i,j \in \{1,2,\ldots, n\}$), $\lim \limits_{t \to \infty} \dot{q} = 0$ and $\lim_{t \to \infty} (b_i-\hat{b}_i^k) = 0$ (for all $i,k \in \{1,2,\ldots, n\}$) exponentially (beyond the collective initial excitation window, i.e. $t \geqslant \bar{T}$) for sufficiently large $\mu_{IF} > 0$, while ensuring that the trajectories of the closed-loop system given by \eqref{stilde_dot}, \eqref{wfull}, and \eqref{ac_law2} are uniformly bounded.
\end{thm}

\begin{rem}
    While the result stated in \Cref{thm:main} pertains to the consensus problem, the same idea extends to the trajectory tracking problems for a known bounded, smooth trajectory $r(t)$ known to the agents. The error variable $e \triangleq (q - r)$ is used in the results and control design instead of $q$.
\end{rem}

\begin{proof}
The closed-loop system using \eqref{eq:clsys}, \eqref{ac_law2}, and \eqref{prop1} can be written in the following matrix structure,
\begin{align*}
    \small{\frac{\dif}{\dif t}\begin{pmatrix} 
                            s \\
                            \tilde{\theta}
                        \end{pmatrix}} &= \small{\begin{pmatrix}
                                            -\sigma I_{mn} & Z_{new}(\dot{q},t) \\
                                            0 & -\mu_F \phi_F-\mu_{IF} \phi_{IF}-\mathcal{L}\otimes I_{(mn+2)}
                                        \end{pmatrix}}\\ & \times\small{\begin{pmatrix}
                                                        \bar{\Lap}(t)s \\
                                                        \tilde{\theta}
                                                    \end{pmatrix}}
\end{align*}
We now define a new consensus error variable $\epsilon \triangleq (I_{mn} - \frac{\one_n\one_n^\top}{n} \otimes I_m )s = (s - \sum_{i=1}^n s_i/n)$. The dynamics in the new error state variables are,
\begin{align}\label{eq:deltadyn}
   &\dot{\epsilon} = -\sigma \Big( \bar{\Lap}(t) + \frac{\one_n\one_n^\top}{n} \otimes I_m  \Big) \epsilon \nonumber\\
   &+ \Big(I_{mn} - \frac{\one_n\one_n^\top}{n} \otimes I_m \Big)Z_{new}(\dot{q},t) \tilde{\theta} \nonumber\\
   & \dot{\tilde{\theta}} = -(\mu_F \phi_F+\mu_{IF} \phi_{IF}+\mathcal{L}\otimes I_{(mn+2)}) \tilde{\theta} 
%   - Z^T(q,\dot{q},t) \bar{\Lap}(t)\epsilon
\end{align}
Arriving at the first equation is a straightforward application of the definition of $\epsilon$, computing its derivative according to the preceding dynamics and noting the fact that $\sigma \bar{\Lap}(t)s = \sigma \bar{\Lap}(t) (\epsilon + \frac{\one_n\one_n^\top}{n} \otimes I_m s)$; $(\frac{\one_n\one_n^\top}{n} \otimes I_m)(I_{mn} - \frac{\one_n\one_n^\top}{n} \otimes I_m) = 0$. Similar transformation equations appear in consensus analysis in \cite[section IIA]{AST17}. 

By applying \Cref{thm:con-pe} we have,
\begin{align*}
    & \Big( \bar{\Lap}(t) +  \frac{\one_n\one_n^\top}{n}\otimes I_m \Big) = (N(t)\otimes I_m)(N(t) \otimes I_m)^\top
\end{align*}
where $N(t) \triangleq \Big(\mathcal{H}(t) + \frac{\one_n h^\top(t)}{\sqrt{n}}\Big)$. It is also evident from \Cref{thm:con-pe} that $N(t)$ is persistently exciting since $\Gr(t)$ is jointly $(\delta,T)$-connected. For the error dynamics \eqref{eq:deltadyn} we choose the following candidate Lyapunov function motivated by \Cref{thm:peV},
\begin{align*}
     & V(t,\epsilon, \tilde{\theta}) = \frac{1}{2}\epsilon^\top [\pi I_{mn} + S(t)]\epsilon + \frac{1}{2} \tilde{\theta}^\top \tilde{\theta}
\end{align*}
where $S(t)$ and constants $\pi, \delta_T$ are as defined in \Cref{thm:peV} with $N(t)$ defined above. It is immediately evident from the definition of $S(t)$ and $\delta_T$ that,
\begin{align*}
    0 \leqslant S(t) \leqslant 2\delta_T I_{mn}
\end{align*}
and the derivative of $S(t)$ can be computed as,
\begin{align*}
    & \dot{S}(t) = 2N(t)N^\top(t) -\frac{2}{T}\int_t^{t+T} N(\tau)N^\top(\tau) \dif \tau. 
\end{align*}
It is therefore obvious that $V(t,\epsilon,\tilde{\theta}) \geqslant 0.5(\pi \| \epsilon \|^2 + \| \tilde{\theta} \|^2)$ and therefore is a positive definite function. The directional derivative of $V(t,\epsilon,\tilde{\theta})$ along the dynamics \eqref{eq:deltadyn} can now be computed as,
\begin{align*}
    &  \dot{V}(t,\epsilon,\tilde{\theta}) = -(\pi \sigma - 1) \epsilon^\top N(t)N^\top(t) \epsilon - \sigma \epsilon^\top S(t)N(t)N^\top(t) \epsilon \\
    & - \frac{1}{T}\epsilon^\top \int_t^{t+T} N(\tau)N^\top(\tau) \dif \tau \epsilon\\
    &+ \epsilon^\top[\pi I_{mn} + S(t)]\bar{I}_{mn}Z_{new}(\dot{q},t) \tilde{\theta} \\
    & -\tilde{\theta}^\top (\mu_F \phi_F+\mu_{IF} \phi_{IF}+\mathcal{L}\otimes I_{(mn+2)}) \tilde{\theta} 
    % - \tilde{\theta}^\top Z^T(q,\dot{q},t) \bar{\Lap}(t)\epsilon
\end{align*}
where $\bar{I}_{mn} \triangleq \Big(I_{mn} -  \frac{\one_n\one_n^\top}{n} \otimes I_m \Big)$ is used as a placeholder. We now compute an upper bound for $\dot{V}(t,\epsilon,\tilde{\theta})$ as below. Keeping in mind that $\phi_F \geqslant 0$, $\phi_{IF} \geqslant 0$ and Proposition \ref{garglemma} we obtain,
\begin{align*}
    & \dot{V}(t,\epsilon,\tilde{\theta}) \leqslant  -(\pi \sigma - 1)\|N^\top(t) \epsilon\|^2 - \frac{\mu_1}{T} \|\epsilon\|^2 + \frac{\sigma \gamma}{2}\|N^\top(t) \epsilon \|^2 \\
    & +\frac{\sigma}{2\gamma}\|S(t)N(t)\|^2\|\epsilon\|^2-\mu_{IF} \lambda_{min}(M(t))\|\tilde{\theta}\|^2\\
    &+ \|[\pi I_{mn} + S(t)]\|\|\bar{I}_{mn}\|\|Z_{new}(\dot{q},t)\| \|\epsilon\|\|\tilde{\theta}\| \\
      \\
    %+ \|Z(q,\dot{q},t)\|\|\bar{\Lap}(t)\|\|\epsilon\|\|\tilde{\theta}\|
   & \leqslant -(\pi \sigma - 1-\frac{\sigma \gamma}{2})\|N^\top(t) \epsilon\|^2 - \big(\frac{\mu_1}{T}- \frac{\sigma}{2\gamma}\|N\|_\infty^2\|S\|_\infty^2\big) \|\epsilon\|^2 \\
   & + z_M(\pi + \|S\|_\infty)\|\bar{I}_{mn}\| \|\epsilon\|\|\tilde{\theta}\|-\mu_{IF} \lambda_{min}(M(t))\|\tilde{\theta}\|^2
\end{align*}
where the first inequality is based on norm upper bounding, utilizing the persistence condition on $N(t)$ (\Cref{def:PE}) and applying the Young's inequality to bound mixed terms in $\epsilon$ using a constant $\gamma > 0$.  We now note that $\|S\|_\infty \leqslant 2\delta_T$ and make the following choice of constants in the above inequality,
\begin{align*}
    & \gamma = \frac{4\sigma T\delta_T^2}{\mu_1} \|N\|_\infty^2; \quad \pi = \frac{1}{\sigma} + \frac{2\sigma T\delta_T^2}{\mu_1} \|N\|_\infty^2
\end{align*}
which leads to,
\begin{align*}
     \dot{V}(t,\epsilon,\tilde{\theta}) &\leqslant -\frac{\mu_1}{2T}\|\epsilon\|^2 + z_M(\pi + 2\delta_T)\|\bar{I}_{mn}\| \|\epsilon\|\|\tilde{\theta}\|\\
    &-\mu_{IF} \lambda_{min}(M(t))\|\tilde{\theta}\|^2 \\
    & \leqslant -\big( \frac{\mu_1}{2T} - \frac{\beta}{2\gamma^o}\big)\|\epsilon\|^2\\
    &- \big(\mu_{IF} \lambda_{min}(M(t)) - \frac{\beta \gamma^o}{2}\big)\|\tilde{\theta}\|^2
\end{align*}
where the final inequality is an application of the Young's inequality with some $\gamma^o > 0$ and $\beta \triangleq z_M(\pi + 2\delta_T)\|\bar{I}_{mn}\|$. We note that $M(t)\geq M(t_0+\bar{T})>0$ $\forall$ $t\geq t_0+\bar{T}$, which implies $\exists$ $c>0$ such that $\lambda_{min}(M(t))\geq c >0$ using the argument as in \Cref{fact_nondecreasing}. Therefore, the following choice of constants guarantees exponential convergence of the $(\epsilon,\tilde{\theta})$ dynamics to the origin.
\begin{align*}
    & \gamma^o > \frac{\beta T}{\mu_1}; \quad \mu_{IF} > \frac{\beta^2 T}{2\mu_1 \lambda_{min}(M(t))}
\end{align*}
We can now argue from the convergence of $\epsilon$ that $\bar{\Lap}(t)s \rightarrow 0$ exponentially. Now employing the definition of $s_i$ in \eqref{s_i} and accounting for the fact that $s_i - s_j \rightarrow 0$ for all $i,j \in \{1,2,\ldots,n\}$ and $\tilde{b}^i \rightarrow 0$ exponentially, we are left with,
\begin{align*}
    \frac{\dif}{\dif t}(q_i - q_j) = -\lambda (q_i - q_j) + \Upsilon(t)
\end{align*}
where $\Upsilon:\R^+ \rightarrow \R^m$ denotes an exponentially decaying function. This immediately shows that $\lim_{t\rightarrow \infty} (q_i - q_j) = 0$ and the convergence is exponential. This implies that $(\bar{\Lap}(t)q, \; \bar{\Lap}(t)\dot{q}) \rightarrow (0,0)$ exponentially, from the above equation. We use these facts to carry out an asymptotic analysis of the closed-loop by substituting $w$ from \eqref{wfull} in \eqref{new_dyn}. Since, $\tilde{b}, \dot{\tilde{b}} \rightarrow 0$, we have the dynamics, in the limit, as $\ddot{q} = -\lambda \dot{q}$, which immediately proves that $\lim_{t\rightarrow \infty} \dot{q} = 0$ exponentially using similar arguments as before.\\
For proof of boundedness, we note that in the $\tilde{\theta}$ dynamics of \eqref{eq:deltadyn}, $\mu_F, \mu_{IF} > 0$ and $\phi_F$, $\phi_{IF}$, $\mathcal{L}\otimes I_{(mn+2)}$ are symmetric positive semidefinite matrices at each $t \geqslant 0$. Therefore we immediately have $\|\tilde{\theta}\| \leqslant \|\tilde{\theta}(0)\|$. It is already known that the unforced ($\tilde{\theta} = 0$) dynamics of $\epsilon$ is exponentially stable (\cite{RSML-18,AST17}) and from the fact that the forcing term is bounded, we can conclude boundedness of $\| \epsilon\|$ irrespective of the collective initial excitation on $Y_{F_i}$. Therefore $\|\bar{\Lap}(t)s\| = \|\bar{\Lap}(t) \epsilon\|$ is also uniformly bounded. Therefore, employing the definition of $s_i$ in \eqref{s_i} we obtain,
\begin{align*}
    & s_i - s_j = \dot{q}_i - \dot{q}_j + \lambda(q_i - q_j) = \psi(t)
\end{align*}
where $\psi:\R^+ \rightarrow R^{mn}$ is a uniformly bounded function ($\|\psi(t)\| \leqslant \psi_M$). We have used the boundedness of $\tilde{b}$ to arrive at the above. Solving the above equation allows us to conclude that $\|(q_i-q_j)\|$ and $\|(\dot{q}_i-\dot{q}_j)\|$ are uniformly bounded.
% The boundedness of the closed-loop trajectories from $[0,\bar{T}]$ is ensured by the Lipschitz control law (preventing finite-escape) and for $t \geqslant \bar{T}$ by the above Lyapunov analysis.
\end{proof}
\begin{rem}
    We note that the first equation in the system \eqref{eq:deltadyn} is identical to the consensus dynamics studied in \cite{RSML-18} and similar in structure to \cite[section IIA]{AST17} if the forcing term due to $\tilde{\theta}$ vanishes. The $\tilde{\theta}$ term is a result of the unknown sensor bias being studied in this article and evolves according to network properties embedded in $\phi_F$, $\phi_{IF}$, and $Z_{new}(\dot{q},t)$.
\end{rem}

\section{Choosing gains and C-IE condition on regressors}
A central condition for exponential convergence of the bias estimation error to zero is \Cref{ass_cie}. We have, $Y_{F_i}(t) = e^{-\beta t}\int_{0}^{t}e^{\beta s}Y_i( \dot{q}_i(\tau) ,\ddot{q}_i(\tau),\tau) \dif \tau$, where from \eqref{Yi}, $Y_i(\dot{q}_i ,\ddot{q}_i,t)$ = $\Big[ \ddot{q}_i, \,\, k(t)\dot{q}_i, \,\, \frac{1}{2}k(t)[Q_{i1}(t)I_m, Q_{i2}(t)I_m,..., Q_{in}(t)I_m]\Big]$. For \Cref{ass_cie} to be satisfied, it is required that the integral of $\sum_{i=1}^n Y_{F_i}^{\top}Y_{F_i}$ over the initial finite time window spans the $mn+2$ dimensional space. We will now prove that, if the set of regressors $Y_i$'s are C-IE then the set of filtered regressors, $Y_{F_i}$'s, are also C-IE which further implies that $Y_{{IF}_i}$'s are C-IE.
%\hlt{\textbf{To show:-} If the set of regressors $Y_i$'s is C-IE then the set of filtered regressors, $Y_{F_i}$'s, is also C-IE which further implies that $Y_{{IF}_i}$'s is C-IE.}
\begin{prop}
The sufficient condition for the set of $Y_{F_i}$'s ($i \in \{1,2,\ldots,n\}$) to be C-IE is that the set of $Y_i$'s are C-IE.
\end{prop}
\begin{proof}
We proceed along the line of proof given in \cite[Proposition 4.1]{Roy19}.
Consider an arbitrary unit vector $v \in \mathbb{R}^{(mn+2)}$ and define the following variables:-
\begin{align*}
    K_i\triangleq Y_iv\\
    K_{F_i}\triangleq Y_{F_i}v
\end{align*}
Let us assume that the set of regressors $Y_i$'s are C-IE.
%\begin{equation}
 %   \int_{t_0}^{t_0+T}\sum_{i=1}^n(K_i^{\top}({\tau})K_i(\tau))d\tau \geq \alpha, \quad \forall v\in \mathbb{R}^{(mn+2)}, \|v\|=1
%\end{equation}
%which implies $\exists$ at least one $i$ for which $K_i(t)\neq 0$ $\forall t \in [t_0, t_0+T]$.\\
The above proposition can now be proved by contradiction. Suppose that $Y_{F_i}$'s are not C-IE. Then, $\exists$ $v \in \mathbb{R}^{(mn+2)}$ such that
\begin{equation*}
  \int_{t_0}^{t_0+T}\sum_{i=1}^n(K_{F_i}^{\top}({\tau})K_{F_i}(\tau))d\tau   = 0
\end{equation*}
which implies that, $K_{F_i}(t)=0$, $\forall i$, $\forall t \in [t_0,t_0+T]$. Therefore, $\dot{K}_{F_i}(t)=0$ $\forall i$ and $t \in (t_0,t_0+T)$. By definition, we have,
\begin{equation*}
    \dot{K}_{F_i}=-\beta K_{F_i}+K_{i}
\end{equation*}
which indicates that $K_{i}(t)$ is zero for all $i$, $\forall t \in (t_0,t_0+T)$. This contradicts the fact that $Y_i$'s are C-IE. Hence, the
set of $Y_i$'s being C-IE implies that the set of $Y_{F_i}$'s are C-IE.
\end{proof}
We now derive a necessary condition to be able to conclude collective Initial Excitation (C-IE) on the set of regressors, $Y_i(\dot{q} ,\ddot{q},t)$. Since, $Y=[Y_1,Y_2,..,Y_n]^{\top}$ we can write,
\begin{align*}%\label{eq:YtY}
\sum_{i=1}^nY_i^\top Y_i(\dot{q}_i ,\ddot{q}_i,t)= & Y^\top Y(\dot{q} ,\ddot{q},t) = \begin{pmatrix}
                        A(\dot{q} ,\ddot{q},t) & B(\dot{q} ,\ddot{q},t) \\
                        B^\top(\dot{q} ,\ddot{q},t) & C(\dot{q},t)
                    \end{pmatrix}, \nonumber\\ \textrm{ where, }
    & A(\dot{q} ,\ddot{q},t) = \nonumber  \begin{pmatrix}
            \ddot{q}^\top \ddot{q} & k(t)\ddot{q}^\top\dot{q} \\
            k(t)\dot{q}^\top \ddot{q} & k^2(t) \|\dot{q}\|^2
         \end{pmatrix}, \nonumber \\
    & B(\dot{q} ,\ddot{q},t) = \begin{pmatrix}
            \frac{k(t)}{2}\ddot{q}^\top \bar{\Lapu}(t) \\
            \frac{k^2(t)}{2}\dot{q}^\top  \bar{\Lapu}(t)
          \end{pmatrix}, \nonumber \\
    & C(t) = \frac{k^2(t)}{4} \bar{\Lapu}(t)^{\top}\bar{\Lapu}(t). \nonumber 
\end{align*}
All arguments in the preceding equation have been deliberately removed for the sake of brevity and clarity.
The functions $A(\dot{q} ,\ddot{q},t)$ and $C(t)$ as defined above, map into positive semidefinite matrices by definition. We now state the result pertinent to this section.

\begin{lem}
\label{thm:necIE}
If $Y_i(\dot{q}_i ,\ddot{q}_i,t)$'s are collectively initially exciting as per \Cref{def:C-IE}, then the matrix functions, $\int_0^{\bar{T}} A(\dot{q} ,\ddot{q},t) \dif t$ and $\int_0^{\bar{T}} C(t) \dif t$ are positive definite. Further, if the graph, $\Gr(t)$ is jointly $(\delta,T)$-connected for some $\delta >0$, then it is also jointly non-bipartite over $[0,\max\{T,\bar{T}\}]$.
\end{lem}
\begin{proof}
Let us assume for contradiction that $0 \in \text{spec}\{\int_0^{\bar{T}} C(t) \dif t\}$. Let the eigenvalues be ordered as $0 \leqslant \beta_2 \leqslant \cdots \leqslant \beta_{mn}$. We already know that $\int_0^{\bar{T}} \sum_{i=1}^{n}Y_i^{\top} Y_i(\dot{q}_i ,\ddot{q}_i,\tau) \dif \tau \geqslant 0$ which indicates that all eigenvalues are non-negative. Let us assume these are ordered as $\lambda_1 \leqslant \lambda_2 \leqslant \cdots \leqslant \lambda_{(mn+2)}$. Since $\int_0^{\bar{T}} C(t) \dif t$ is a principal submatrix of $\int_0^{\bar{T}}\sum_{i=1}^n Y_i^{\top} Y_i(\dot{q}_i ,\ddot{q}_i,\tau) \dif \tau$, we can use the Cauchy's interlacing and inclusion theorem (\cite[Theorem 8.4.5]{DB-09}) to conclude that, $\lambda_1 \leqslant 0 \leqslant \lambda_3$. Since $\int_0^{\bar{T}}\sum_{i=1}^n Y_i^{\top} Y_i(\dot{q}_i ,\ddot{q}_i,\tau) \dif \tau \geqslant 0$, the only possibility is $\lambda_1 = 0$. This immediately implies that the set of $Y_i(\dot{q}_i ,\ddot{q}_i,t)$'s are \emph{not} collectively initially exciting, thus contradicting our premise. Similar arguments can be used to claim that $\int_0^{\bar{T}} A(\dot{q} ,\ddot{q},t) \dif t > 0$. \\
From $\int_0^{\bar{T}} C(t) \dif t > 0$ along with the facts that $k(t)>0$, we can immediately conclude $\int_0^{\bar{T}} \bar{\Lapu}^2(t) \dif t \geqslant \lambda_c I_{mn}$ for some $\lambda_c > 0$. This immediately implies that $\int_0^{\max\{T,\bar{T}\}} \bar{\Lapu}(t) \dif t > 0$. The union graph of $\Gr(t)$, denoted $\cup_{t \in [0,\max\{T,\bar{T}\}]} \Gr(t)$, is connected by the joint $(\delta,T)$-connectedness assumption. The positive definiteness of the signless Laplacian for the union graph ($\int_0^{\max\{T,\bar{T}\}} \bar{\Lapu}(t) \dif t$) and the joint connectivity of the union graph allows us to invoke \Cref{thm:bip} and conclude that the union graph $\cup_{t \in [0,\max\{T,\bar{T}\}]} \Gr(t)$ is non-bipartite, i.e. $\Gr(t)$ is jointly non-bipartite over $[0,\max\{T,\bar{T}\}]$.
\end{proof}

\begin{rem}
    We note here that \cite{SDTM-20} utilizes the non-bipartite graph structure to propose exponential bias estimators. The graph is, however, assumed to be constant. The necessary condition above allows the use of time-varying graphs that are only jointly non-bipartite (as opposed to at each time instant) and persists only for a finite time $[0,\max\{T,\bar{T}\}]$.
\end{rem}

Based on \Cref{thm:necIE}, assuming that we have a jointly non-bipartite union graph over $[0,\max\{T,\bar{T}\}]$, the primary purpose of $k(t)$ is to ensure that $\int_0^{\bar{T}} A(\dot{q} ,\ddot{q},t) \dif t$ becomes positive definite. While there is no direct way to prescribe such a function for all possible initial conditions and system parameters, we introduce multiple frequency components through the time dependence in $k(t)$ to make the columns of $A(\dot{q} ,\ddot{q},t)$ linearly independent over sub-intervals.

\section{Simulation Results}
% unfortunately this example may land us in some trouble, for example it is not clear what kind of sensors can be used to measure relative angular positions between agents and how bias shows up there. In linear measurements, like the spacecraft case things are much more clear. If you look below, I have added a reference [20] for the same. Also you trajectory to track/leader is rather high in frequency and magnitude, you should use data from our SC 617 course project.
We now present simulation studies to verify \Cref{thm:main} for a network of double integrators interacting via an undirected graph $\Gr(t)$ and bias corrupted measurements. We consider the translational dynamics of $n$ identical quadrotors given by
\begin{align}\label{positionquad}
    \ddot{q}_i=\begin{bmatrix}0 \\0\\ -9.8\end{bmatrix}+\textbf{R}_ie_3\frac{\tau_{1_i}}{M},\quad \forall i=1,2,..,n
\end{align}
where the position vector is denoted by $q_i=(x_i,y_i,z_i)^{\top} \in \mathbb{R}^3$, $e_3=(0,0,1)^{\top}$, $M\in \mathbb{R}^+$ is the mass of the quadrotor. $\textbf{R}_i \in SO(3)$ is the $3 \times 3$ orthogonal rotation matrix from the quadrotor body frame to the inertial frame. The feedback $\tau_{1_i}(\cdot) \in \mathbb{R}$ is the sum of thrust forces from the individual motors in each quadrotor. Typical tracking control of the quadrotor consists of an inner loop attitude control \cite{sridharfinite} which modulates $\textbf{R}_i$ while the outer loop translation control is designed assuming full linear actuation in (\ref{positionquad}). Therefore, (\ref{positionquad}) can be treated as a double integrator model for our purposes by assuming a new control $u_i := \textbf{R}_ie_3\frac{\tau_{1_i}}{M}$. For these simulations we have considered $n=5.$

\begin{figure}[!htbp]
   %\begin{minipage}{0.22\textwidth}
   \begin{minipage}{0.45\columnwidth}
     \centering
     \includegraphics[scale=0.4]{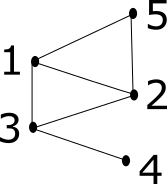}
     \caption{Connected non-Bipartite graph}\label{test1}
   \end{minipage}\hfill
   \begin{minipage}{0.45\columnwidth}
     \centering
     \includegraphics[scale=0.4]{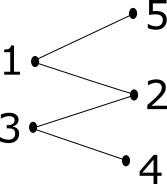}
     \caption{Connected Bipartite graph}\label{test2}
   \end{minipage}
\end{figure}

%For these simulations $n=5$, $m=2$ and the state variables are distances denoted by $q_i = (q_i^1,q_i^2)^{\top}$. The relative distances between neighbors are assumed to be measured.
Two adjacency matrices are used, one corresponding to a non-bipartite, connected graph($\mathcal{A}_b$) and another corresponding to a bipartite, connected graph($\mathcal{A}_c$).
%\begin{align*}
            %&M_i(q_i) = \begin{bmatrix} 
           % p_1 + 2p_3cos(q_i^2) & p_2+p_3cos(q_i^2)\\
           % p_2+p_3cos(q_i^2) & p_2
            %\end{bmatrix},  \\
            %&C_i(q_i,\dot{q}_i) = \begin{bmatrix}
            %-p_3sin(q_i^2)\dot{q_i^2} & %-p_3sin(q_i^2)(\dot{q}_i^1+\dot{q}_i^2)\\
           % p_3sin(q_i^2)\dot{q}_i^1 & 0
            %\end{bmatrix}, \\
            %&F_i(\dot{q}_i) = \begin{bmatrix}
            %f_{d_1} & 0\\ 0 & f_{d_2}
            %\end{bmatrix}, \quad  g_i(q) = [0,\;0]^\top
%\end{align*}
% The dynamics of the agents is given by,
% \begin{align}
%     \label{EL_sim}
%     M_i(q)_i\ddot{q}_i + C_i(q_i,\dot{q}_i)\dot{q}_i + F_i(\dot{q}_i)\dot{q}_i = \tau_i
% \end{align}
% with $q_{i1},\,q_{i2}$ being the joint angle for the two joints and $\tau_i$ representing the external torques applied to the revolute joints. The gravity component is not present in \eqref{EL_sim} since we are considering planar robot manipulators. 
%where $p_1 = 3.31, \; p_2 =  0.116, \; p_3 = 0.16, \;f_{d_1} = 5.3, \;f_{d_2} = 1.1$. 
% The leader's trajectory is given by $q_0 = [0.04\mathtext{sin}(0.25t) \, ; \, 0.04\mathtext{sin(0.25t)}]$. 
%The relative angles between neighbors are assumed to be measured using optical or capacitive relative angle sensors~\cite{fwzbc02}. Two adjacency matrices are used, one corresponding to a non-Bipartite, connected graph ($\Adj_b$) and another corresponding to a Bipartite, connected graph ($\Adj_c$).
\begin{align*}
    &\Adj_b = \begin{bmatrix}
    0 & 1 & 1 & 0 & 1\\
    1 & 0 & 1 & 0 & 1\\
    1 & 1 & 0 & 1 & 0\\
    0 & 0 & 1 & 0 & 0\\
    1 & 1 & 0 & 0 & 0\\
    \end{bmatrix}, \quad
    %\end{align*}}
    %\hlt{\begin{align*}
    \Adj_c = \begin{bmatrix}
    0 & 1 & 0 & 0 & 1 \\
    1 & 0 & 1 & 0 & 0 \\
    0 & 1 & 0 & 1 & 0 \\
    0 & 0 & 1 & 0 & 0 \\
    1 & 0 & 0 & 0 & 0 \\
    \end{bmatrix}.
\end{align*}
The graphs corresponding to $\mathcal{A}_b$ and $\mathcal{A}_c$ are shown in \Cref{test1} and \Cref{test2}, respectively.

\begin{figure}[!htbp]
\centering
\includegraphics[width=\columnwidth,height=0.7\columnwidth]{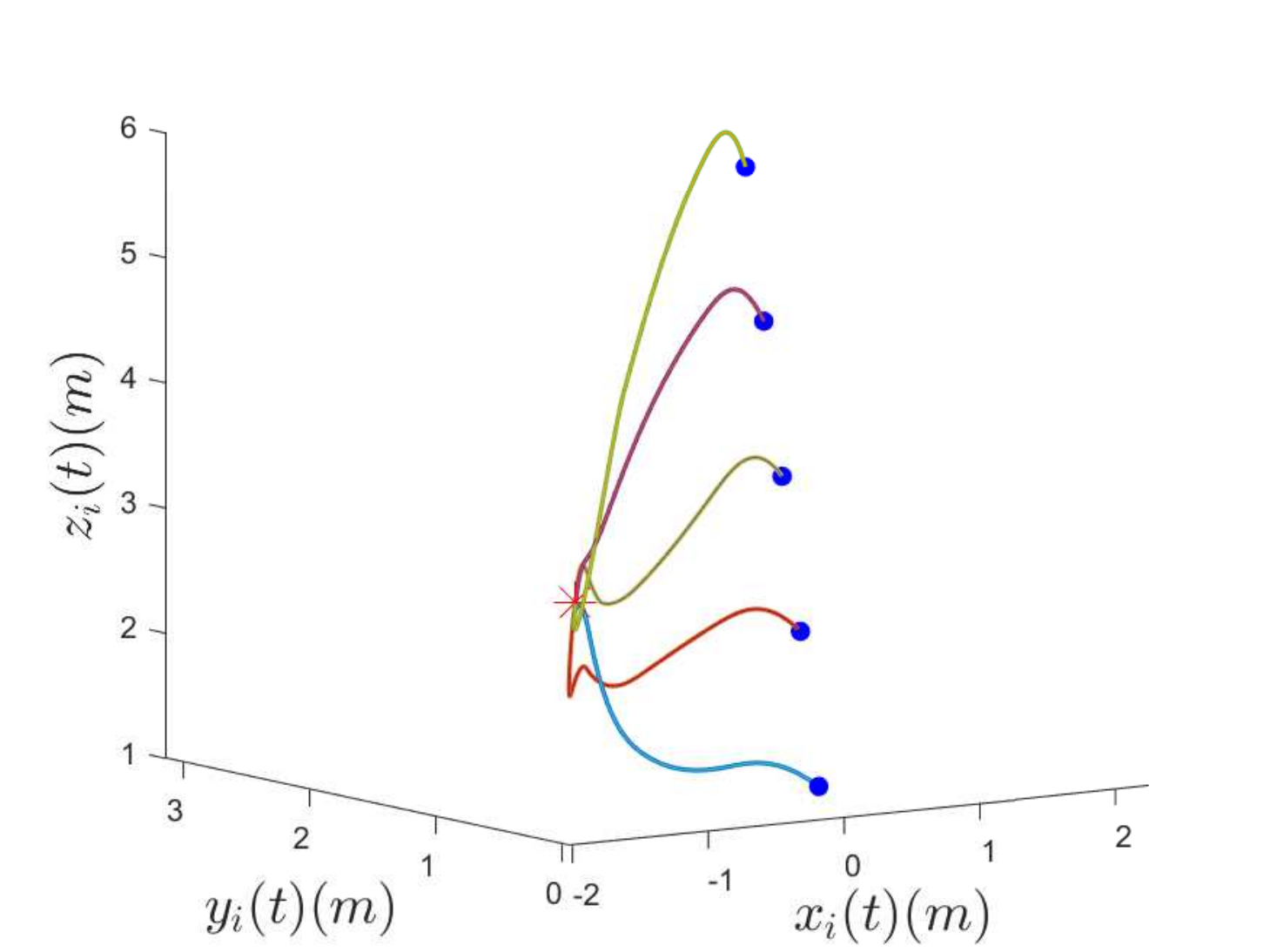}
\caption{$x_i(t)(m)$ vs $y_i(t)(m)$ vs $z_i(t)(m)$. The blue dots indicate start locations, and the red asterisk the terminal locations.}
\label{pos_fig}       % Give a unique label
\end{figure}

The actual adjacency matrix is obtained by cycling periodically (period $T=8s$) between sub graphs of $\mathcal{A}_b$ for the first $t=8$ seconds and then between sub graphs of $\mathcal{A}_c$ for the rest of the time. This allows for a jointly non-bipartite and connected graph in the initial phase of the simulations and a jointly bipartite, connected graph beyond.
\par The initial position, velocity, and the bias in relative measurement of the position for the $i^{th}$ quadrotor are given by [$\frac{i\pi}{7}$; $\frac{i\pi}{5}$; $\frac{i\pi}{3}$], [$0.1i-0.7$; $-0.1i+0.6$; $0.1i+0.7$ ], [$\frac{i \pi}{12}$; $\frac{i \pi}{12}$; $\frac{i \pi}{12}$] respectively for $i=1,2,..,5$. $\hat{\theta}^i$ is initialized to the zero vector for $i=1,2,..,5$. The gain constants $\sigma$, $\mu_{F}$, $\mu_{IF}$, $\lambda$, and $\beta$ are chosen to be 0.2, 0.020, 15, 0.5, 0.5 respectively and $k(t)=1+0.5 cos^2(t)+0.5 sin^2(2t)$. The gain chosen helps introduce multiple frequency components through the time dependence in $k(t)$. The chosen $k(t)$ ensures that $\int_0^{\bar{T}} A(\dot{q} ,\ddot{q},t) \dif t$ becomes positive definite, which guarantees collective initial excitation on the regressor $Y_i$'s.
\begin{figure}[!htbp]
\centering

\includegraphics[width=\columnwidth,height=0.7\columnwidth]{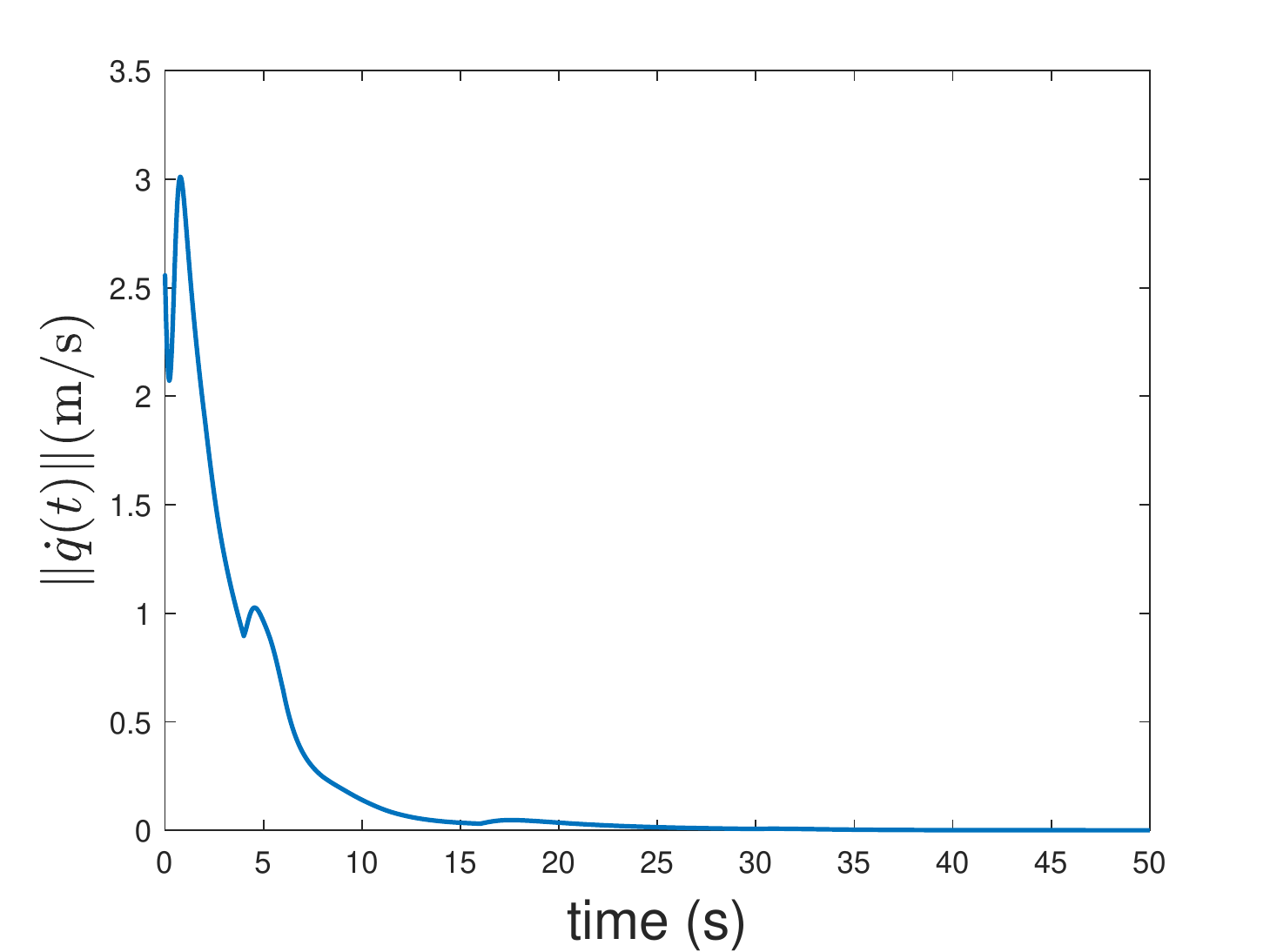}
\caption{$\|\dot{q}(t)\|(m/s)$ vs time ($s$).}
\label{vel1_fig}       % Give a unique label
\end{figure}

%\begin{figure}[!htbp]
%\centering
% \sidecaption
%\includegraphics[width=7.0cm,height=5.0cm]{normvlogscale.eps}
%\caption{$\|\dot{q}(t)\|(m/s)$ vs time ($s$). The $y$-axis is in a logarithmic scale.}
%\label{vel_fig}       % Give a unique label
%\end{figure}
 %The initial position, velocity and the bias in relative measurement of the position for the $i^{th}$ agent are given by $[ \frac{i\pi}{7} \, ; \, \frac{i\pi}{8} ]$, $[0.1i-0.7 \, ; \, -0.1i+0.6]$ and $[ \frac{i\pi}{12} \, ; \, \frac{i\pi}{12}]$ respectively for $i = 1,2,\ldots, 5$. $\hat{\theta}$ is initialized to the zero vector. The gain constants $\sigma, \; \mu_F,\;\mu_{IF}, \; \lambda, \; \mathtext{and} \; \beta$ are chosen to be $0,5,\;,0.015,\;15,\;0.4,\;0.1$ respectively and $k (t) = 1 + 0.5\cos^2(t) + 0.5\sin^2(2t)$. 

\begin{figure}[!htbp]
\centering
\includegraphics[width=\columnwidth,height=0.7\columnwidth]{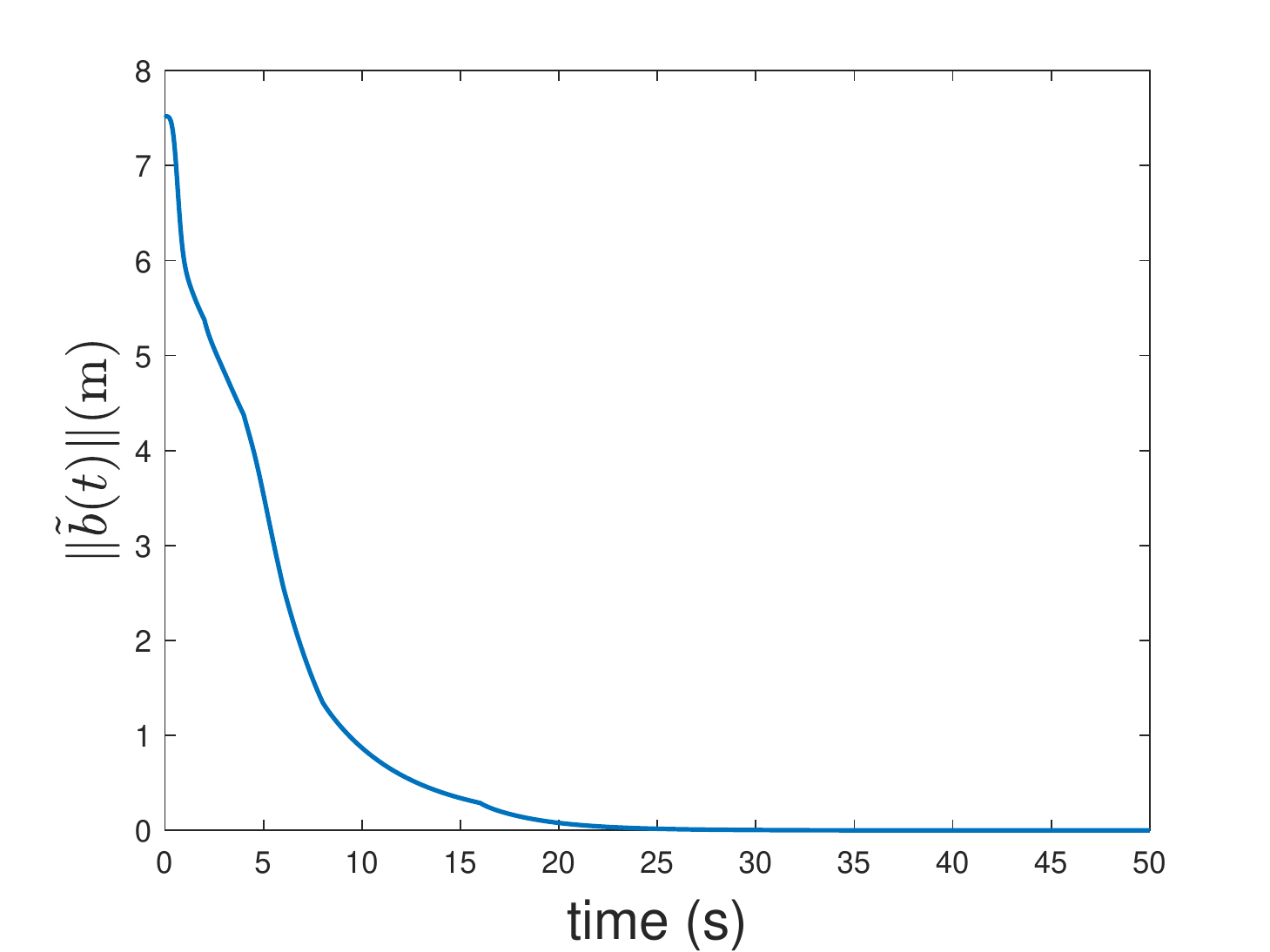}
\caption{$\|\tilde{b}(t)\|(m)$ vs time ($s$).}
\label{btilde1_fig}       % Give a unique label
\end{figure}

\begin{figure}[!htbp]
\centering
\includegraphics[width=\columnwidth,height=0.7\columnwidth]{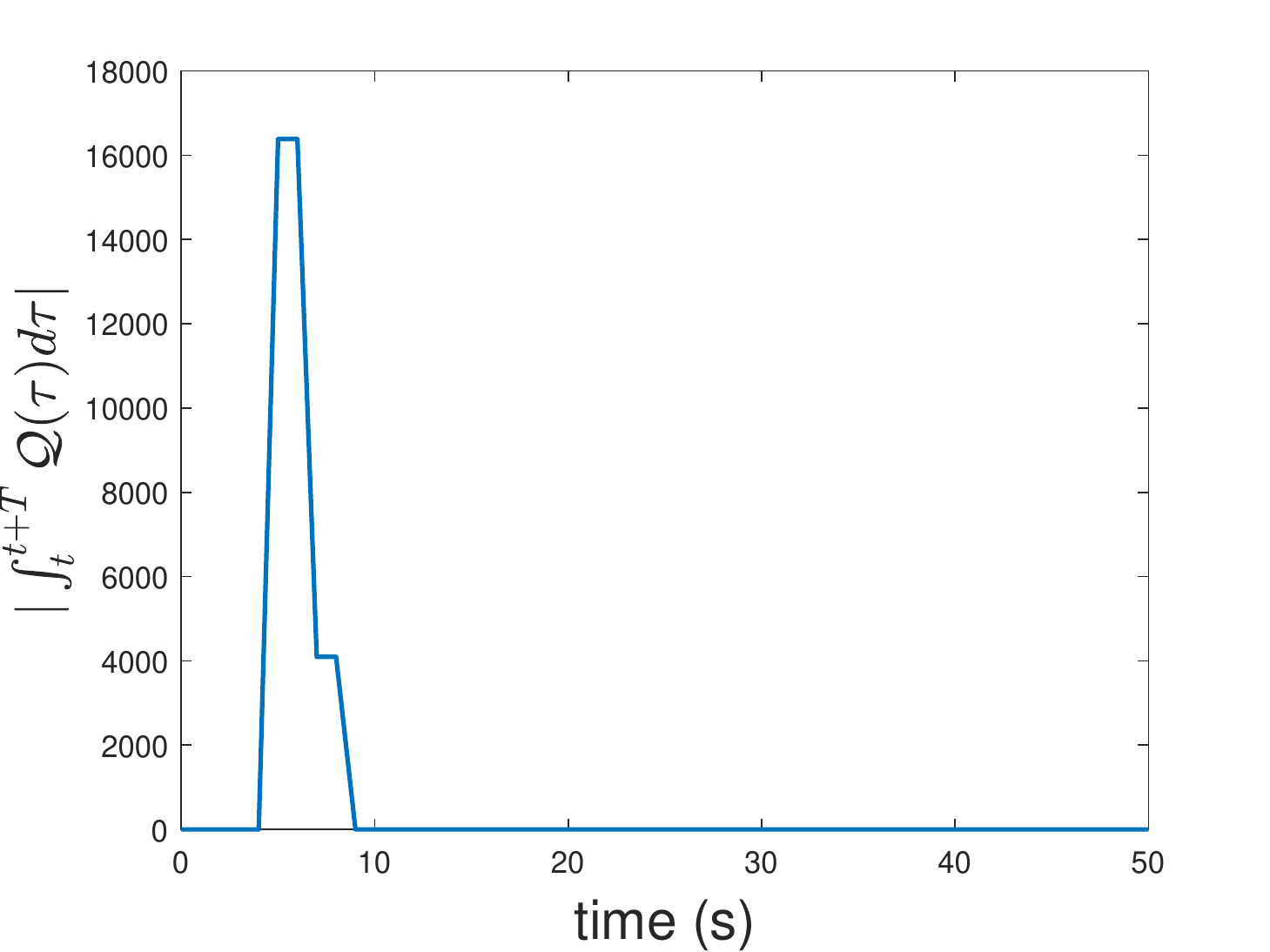}
\caption{Determinant of $\int_t^{t+T}\Lapu(\tau) \dif \tau$ over time $t$, $T = 4 s$.}
\label{control_fig}       % Give a unique label
\end{figure}

 \Cref{pos_fig} is the phase-plane evolution of the three positions. As is evident, starting from different initial conditions, they converge to consensus. Similarly, all the three velocities in \Cref{vel1_fig} and the bias estimation errors given by $\tilde{b}=[\tilde{b}^1,\tilde{b}^2,\tilde{b}^3,\tilde{b}^4,\tilde{b}^5]^{\top}$ in \Cref{btilde1_fig} converge to zero during the simulation horizon. The final plot, \Cref{control_fig} is for the verification of \Cref{thm:necIE}, wherein we claim that the collective initial excitation of the regressors necessitates a jointly non-bipartite graph. \Cref{control_fig} plots the determinant of $\int_t^{t+T}\Lapu(\tau) \dif \tau$ for all $t$, keeping $T=4s$ as the cycling period. We see that the $\int_t^{t+T}\Lapu(\tau) \dif \tau$ is positive definite over an initial period of time and beyond this is singular. This verifies, by \Cref{thm:bip} that $\Gr(t)$ determined by $\Adj(t)$ is jointly non-Bipartite over a finite initial window.

\section{Conclusion}

In this article we propose a novel distributed adaptive controller to estimate bias in relative position measurements along with guaranteed exact consensus in a network of double-integrator systems. It is shown that joint $(\delta,T)$-connectivity and joint non-Bipartite properties of the graph are necessary for bias estimation and consensus. In future work, we seek to explore more general measurement errors and nonlinear agent dynamics. We will also focus on consensus under erroneous relative measurements over directed and time varying communication graphs.
\section{Acknowledgement}
The authors thank Sayan Basu Roy (IIIT Delhi, India), Antonio Loria and Elena Panteley (LSS, CNRS,France), Constantin Morarescu and Vineeth Varma (CRAN, CNRS, France) for their valuable technical inputs.

\bibliographystyle{IEEEtran}
\bibliography{references}

\end{document}